\documentclass[twocolumn]{aastex62}
\pdfoutput=1 

\usepackage{cleveref}
\usepackage{graphicx} 
\usepackage{longtable}
\crefformat{section}{\S#2#1#3} 
\crefformat{subsection}{\S#2#1#3}
\crefformat{subsubsection}{\S#2#1#3}

    \makeatletter
\def\@fnsymbol#1{\ensuremath{\ifcase#1\or \dagger\or \ddagger\or
   \mathsection\or \mathparagraph\or \|\or **\or \dagger\dagger
   \or \ddagger\ddagger \else\@ctrerr\fi}}
    \makeatother

\graphicspath{{./}{figures/}}

\shorttitle{HAT-P-32A\MakeLowercase{b}}
\shortauthors{Alam et al.}

\begin{document}

\title{The \textit{HST} P\MakeLowercase{an}CET Program: An Optical to Infrared Transmission Spectrum of HAT-P-32A\MakeLowercase{b}}

\correspondingauthor{Munazza K. Alam}
\email{munazza.alam@cfa.harvard.edu}

\author[0000-0003-4157-832X]{Munazza K. Alam}
\altaffiliation{National Science Foundation Graduate Research Fellow}
\affiliation{Center for Astrophysics ${\rm \mid}$ Harvard {\rm \&} Smithsonian, 60 Garden Street, Cambridge, MA 01238, USA}

\author[0000-0003-3204-8183]{Mercedes L\'opez-Morales}
\affil{Center for Astrophysics ${\rm \mid}$ Harvard {\rm \&} Smithsonian, 60 Garden Street, Cambridge, MA 01238, USA}

\author[0000-0002-6500-3574]{Nikolay Nikolov}
\affil{Space Telescope Science Institute, 3700 San Martin Dr, Baltimore, MD 21218, USA}

\author[0000-0001-6050-7645]{David K. Sing}
\affil{Department of Physics \& Astronomy, Johns Hopkins University, Baltimore, MD 21218, USA}

\author[0000-0003-4155-8513]{Gregory W. Henry}
\affil{Center of Excellence in Information Systems, Tennessee State University, Nashville, TN  37209,  USA}

\author[0000-0003-3438-843X]{Claire Baxter}
\affil{Anton Pannekoek Institute for Astronomy, University of Amsterdam, Science Park 904, 1098 XH Amsterdam, The Netherlands}

\author[0000-0002-0875-8401]{Jean-Michel D\'esert}
\affil{Anton Pannekoek Institute for Astronomy, University of Amsterdam, Science Park 904, 1098 XH Amsterdam, The Netherlands}

\author[0000-0003-3726-5419]{Joanna K. Barstow}
\affil{School of Physical Sciences, The Open University, Walton Hall, Milton Keynes, MK7 6AA, UK}

\author[0000-0001-5442-1300]{Thomas Mikal-Evans}
\affil{Kavli Institute for Astrophysics and Space Research, Massachusetts Institute of Technology, Cambridge, MA 02139, USA}

\author[0000-0002-9148-034X]{Vincent Bourrier}
\affil{Observatoire de l'Universit\'e de Gen\`eve, Sauverny, Switzerland}

\author[0000-0002-5360-3660]{Panayotis Lavvas}
\affil{Groupe de Spectrom\'etrie Moleculaire et Atmosph\'erique, Universit\'e de Reims Champagne Ardenne, Reims, France}

\author[0000-0003-4328-3867]{Hannah R. Wakeford}
\affil{School of Physics, University of Bristol, HH Wills Physics Laboratory, Tyndall Avenue, Bristol BS8 1TL, UK}

\author{Michael H. Williamson}
\affil{Center of Excellence in Information Systems, Tennessee State University, Nashville, TN  37209,  USA}

\author[0000-0002-1600-7835]{Jorge Sanz-Forcada}
\affil{Centro de Astrobiolog\'ia (CSIC-INTA), ESAC Campus, Villanueva de la Ca\~nada, Madrid, Spain}

\author[0000-0003-1605-5666]{Lars A. Buchhave}
\affil{DTU Space, National Space Institute, Technical University of Denmark, Elektrovej 328, DK-2800 Kgs. Lyngby, Denmark}

\author[0000-0003-3721-0215]{Ofer Cohen}
\affil{Lowell Center for Space Science and Technology, University of Massachusetts, Lowell, MA 01854, USA}

\author[0000-0003-1756-4825]{Antonio Garc\'ia Mu\~noz}
\affil{Technische Universit\"at Berlin EW 801, Hardenbergstra\ss e 36, D-10623 Berlin, Germany}

\begin{abstract}
We present a 0.3$-$5 $\mu$m transmission spectrum of the hot Jupiter HAT-P-32Ab observed with the Space Telescope Imaging Spectrograph (STIS) and Wide Field Camera 3 (WFC3) instruments mounted on the \textit{Hubble Space Telescope}, combined with \textit{Spitzer} Infrared Array Camera (IRAC) photometry. The spectrum is composed of 51 spectrophotometric bins with widths ranging between 150 and 400 \AA, measured to a median precision of 215 ppm. Comparisons of the observed transmission spectrum to a grid of 1D radiative-convective equilibrium models indicate the presence of clouds/hazes, consistent with previous transit observations and secondary eclipse measurements. To provide more robust constraints on the planet's atmospheric properties, we perform the first full optical to infrared retrieval analysis for this planet. The retrieved spectrum is consistent with a limb temperature of 1248$_{-92}^{+92}$ K, a thick cloud deck, enhanced Rayleigh scattering, and $\sim$10x solar H$_{2}$O abundance. We find log($Z/Z_{\odot}$) = 2.41$_{-0.07}^{+0.06}$, in agreement with the mass-metallicity relation derived for the Solar System. 


\end{abstract}


\section{Introduction} \label{sec:intro}

The study of exoplanet atmospheres can provide key insights into planetary formation and evolution, atmospheric structure, chemical composition, and dominant physical processes (\citealt{Seager2010,Crossfield15,Deming17}). Close-in giant planets with extended hydrogen/helium atmospheres are ideal targets for atmospheric characterization via transmission spectroscopy (\citealt{Seager00,Brown01}). The gaseous atmospheres of such targets are accessible from the \textit{Hubble Space Telescope} (\textit{HST}) with the Space Telescope Imaging Spectrograph (STIS) (e.g., \citealt{Charbonneau02,Huitson13,Sing15,Nikolov14,Alam18,Evans18}), and Wide Field Camera 3 (WFC3) (e.g., \citealt{Kreidberg15b,Evans16,Wakeford17,Spake18,Arcangeli18}) instruments. Observational campaigns on large ground-based telescopes (e.g., \citealt{Sing12,Jordan13,Rackham17,Chen17,Louden17,Huitson17,Nikolov18b,Espinoza19,Weaver20}) are also expanding the number of giant planets characterized using this technique. 

Transmission spectra are primarily sensitive to the relative abundances of different absorbing species and the presence of aerosols (e.g., \citealt{Deming18}). Optical transit observations are of particular value because they provide information about condensation clouds and photochemical hazes in exoplanet atmospheres. Rayleigh or Mie scattering produced by such aerosols causes a steep continuum slope at these wavelengths \citep{Lecavelier08}, which can be used to infer cloud composition and to constrain haze particle sizes (e.g., \citealt{Wakeford17b,Evans18}). Combining optical and near-infrared observations can provide constraints on the metallicity of a planet via H$_{2}$O abundance as well as constraints on any cloud opacities present (e.g., \citealt{Wakeford18,Pinhas19}).

We have observed a diversity of cloudy to clear atmospheres for close-in giant planets \citep{Sing16}, but it is currently unknown what system parameters sculpt this diversity. The \textit{HST}/WFC3 1.4 $\mu$m H$_{2}$O feature has been suggested as a near-infrared diagnostic of cloud-free atmospheres correlated with planetary surface gravity and equilibrium temperature \citep{Stevenson16}. The analogous optical cloudiness index of \citet{Heng16} hints that higher temperature (more irradiated) planets may have clearer atmospheres with fewer clouds consisting of sub-micron sized particles. In addition to understanding the physics and chemistry of exoplanet atmospheres, probing trends between the degree of cloudiness in an atmosphere and the properties of the planet and/or host star is important for selecting cloud-free planets for detailed atmospheric follow-up with the \textit{James Webb Space Telescope} (\textit{JWST}). Identifying such targets with current facilities is an important first step.

Optical and near-infrared wavelengths probe different atmospheric layers, so it is possible for one layer to be cloud-free while the other is cloudy. Some planets may be predicted to be cloud-free based on the \citet{Heng16} optical cloudiness index, but not according to the \citet{Stevenson16} near-infrared H$_{2}$O$-$J index. One such planet is the inflated hot Jupiter HAT-P-32Ab ($M_{p}$ = 0.86 $\pm$ 0.16 $M_{J}$; $R_{p}$ = 1.79 $\pm$ 0.03 $R_{J}$; $\rho$ = 0.18 $\pm$ 0.04 g/cm$^{3}$, $T_{eq}$ = 1801 $\pm$ 18 K; $g$ = 6.0 $\pm$ 1.1 m/s$^{2}$), which is the subject of this study.  HAT-P-32Ab is ideal for atmospheric observations with transmission spectroscopy, given its 2.15 day orbital period, large atmospheric scale height (H $\approx$ 1100 km), and bright ($V$ = 11.29 mag) late-type F stellar host \citep{Hartman11}. 


Previous ground-based observations of HAT-P-32Ab's atmosphere reveal a flat, featureless optical transmission spectrum between 0.36 and 1 $\mu$m, consistent with the presence of high altitude clouds \citep{Gibson13,Mallonn16a,Nortmann16}. Short wavelength (0.33$-$1 $\mu$m) broadband spectrophotometry to search for a scattering signature in the blue also yielded a flat transmission spectrum \citep{Mallonn16b}, but near-UV transit photometry in the U-band (0.36 $\mu$m) suggests the presence of magnesium silicate aerosols larger than 0.1 $\mu$m in the atmosphere of HAT-P-32Ab \citep{Mallonn17}. Follow-up high-precision photometry indicates a possible bimodal cloud particle distribution, including gray absorbing cloud particles and Rayleigh-like haze \citep{Tregloan17}. 

In the near-infrared, transit observations reveal a weak water feature at 1.4 $\mu$m, consistent with the presence of high-altitude clouds \citep{Damiano17}. Secondary eclipse measurements of HAT-P-32Ab are consistent with a temperature inversion due to the presence of a high-altitude absorber and inefficient heat redistribution from the dayside to the nightside \citep{Zhao14}. \textit{HST}/WFC3 secondary eclipse measurements from \citet{Nikolov18} find an eclipse spectrum that can be described by a blackbody of $T_{p}$ = 1995 $\pm$ 17 K or a spectrum of modest thermal inversion with an absorber, a dusty cloud deck, or both.

In this paper, we present the optical to infrared transmission spectrum of the hot Jupiter HAT-P-32Ab measured from 0.3$-$5 $\mu$m using the STIS and WFC3 instruments aboard \textit{HST} and the IRAC instrument on \textit{Spitzer}. The STIS observations were obtained as part of the \textit{HST} Panchromatic Comparative Exoplanetology Treasury (PanCET) program (GO 14767; PIs Sing \& L\'opez-Morales). We compare this new broadband spectrum to previous observations of this planet and perform the first optical to infrared retrieval analysis of its atmospheric properties. The structure of the paper is as follows. We describe the observations and data reduction methods in \S \ref{sec:obs_data} and detail the light curve fits in \S \ref{sec:analysis}. In \S \ref{sec:results}, we present the transmission spectrum  compared to previous studies and describe the results from our forward model fits and retrievals. We contextualize HAT-P-32Ab within the broader exoplanet population in \S \ref{sec:context}. The results of this work are summarized in \S \ref{sec:summary}. 

\section{Observations \& Data Reduction} \label{sec:obs_data}

We observed three transits of HAT-P-32Ab with \textit{HST}/STIS (GO 14767, PI: Sing \& L\'opez-Morales) and one transit with \textit{HST}/WFC3 (GO 14260, PI: Deming). Two additional transits were observed with \textit{Spitzer}/IRAC (GO 90092, PI: D\'esert). 

\subsection{HST/STIS}

We obtained time series spectroscopy during two transits of HAT-P-32Ab using \textit{HST}'s Space Telescope Imaging Spectrograph (STIS) on UT 2017 March 6 and UT 2017 March 11 with the G430L grating, which provides low-resolution (R$\sim$500) spectroscopy from 2892$-$5700 \AA. We observed an additional transit with the G750L grating on UT 2017 June 22, which covers the 5240$-$10270 \AA~ wavelength range at R$\sim$500. The visits were scheduled to include the transit event in the third orbit and provide sufficient out-of-transit baseline flux as well as good coverage between second and third contact. Each visit consisted of five consecutive 96-minute orbits, during which 48 stellar spectra were obtained over exposure times of 253 seconds. To decrease the readout times between exposures, we used a 128 pixel wide sub-array. The data were taken with the 52 x 2 arcsec$^{2}$ slit to minimize slit light losses. This narrow slit is small enough to exclude any flux contribution from the M dwarf companion to HAT-P-32A, located $\sim$2.9" away from the target \citep{Zhao14}.   

\begin{figure}
    \centering
    \includegraphics[scale=0.53]{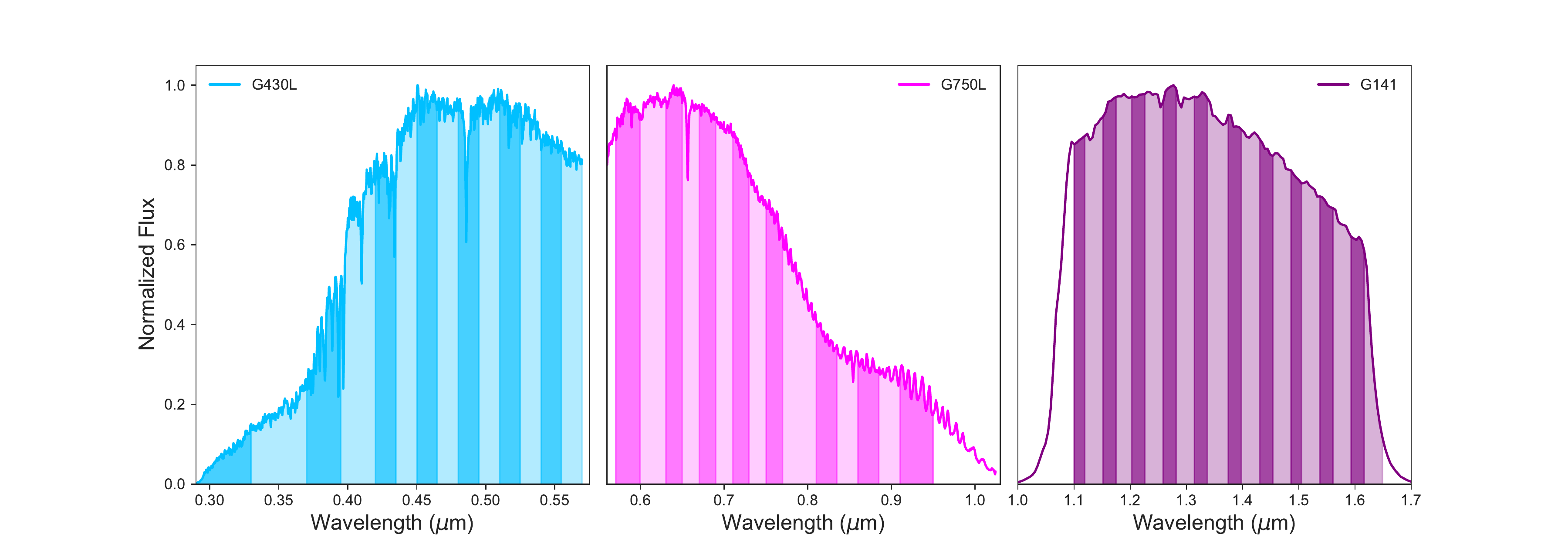}
        \caption{Example stellar spectra for the \textit{HST} STIS G430L (blue; left), STIS G750L (pink; middle), and WFC3 G141 (purple; right) grisms. Vertical bands indicate the wavelength bins adopted for the spectrophotometric light curves (\S \ref{sec:spec_lc}).}
    \label{fig:stellar_spec}
\end{figure}

We reduced the STIS G430L and G750L spectra using the techniques described in \citealt{Nikolov14,Nikolov15} and \citealt{Alam18}, which we summarize briefly here. We used the {\tt CALSTIS} pipeline (version 3.4) to bias-, dark-, and flat-field correct the raw 2D data frames. To identify and correct for cosmic ray events, we used median-combined difference images to flag bad pixels and interpolate over them. We then extracted 1D spectra from the calibrated {\tt .flt} files and extracted light curves using aperture widths of 6 to 18 pixels, with a step size of 1. Based on the lowest photometric dispersion in the out-of-transit baseline flux, we selected an aperture of 13 pixels for use in our analysis. We computed the mid-exposure time in MJD for each exposure. From the {\tt x1d} files, we re-sampled all of the extracted spectra and cross-correlated them to a common rest frame to obtain a wavelength solution. Since the cross-correlation measures the shift of each stellar spectrum with respect to the first spectrum of the time series, we re-sampled the spectra to align them and remove sub-pixel drifts associated with the different locations of the spacecraft on its orbit \citep{Huitson13}. Example spectra for the G430L and G750L gratings are shown in Figure \ref{fig:stellar_spec}.

\subsection{HST/WFC3}
We observed a single transit of HAT-P-32Ab with the Wide Field Camera 3 (WFC3) instrument on UT 2016 January 21. The transit observation consisted of five consecutive \textit{HST} orbits, with 18 spectra taken during each orbit. At the beginning of the first orbit, we took an image of the target using the F139M filter with an exposure time of 29.664 seconds. We then obtained time series spectroscopy with the G141 grism (1.1$-$1.7 $\mu$m). Following standard procedure for WFC3 observations of bright targets (e.g., \citealt{Deming13,Kreidberg14,Evans16,Wakeford17}), we used the spatial scan observing mode to slew the telescope in the spatial direction during an exposure. This technique allows for longer exposures without saturating the detector \citep{McCullough12}. We read out using the  SPARS10 sampling sequence with five non-destructive reads per exposure (NSAMP=5), which resulted in integration times of 89 seconds.   

We started our analysis of the WFC3 spectra using the flat-fielded and bias-subtracted {\tt ima} files produced by the CALWF3 pipeline\footnote{\url{http://www.stsci.edu/hst/wfc3/pipeline/wfc3_pipeline}} (version 3.3). We extracted the flux for each exposure by taking the difference between successive reads and then subtracting the median flux in a box 32 pixels away from the stellar spectrum. This background subtraction technique masks the area surrounding the 2D spectrum to suppress contamination from nearby stars and companions, including the M dwarf companion to HAT-P-32A. We then corrected for cosmic ray events using the method of \citet{Nikolov14}. 

Stellar spectra were extracted by summing the flux within a rectangular aperture centered on the scanned spectrum along the full dispersion axis and along the cross-dispersion direction ranging from 48 to 88 pixels. We determined the wavelength solution by cross-correlating each stellar spectrum to a grid of simulated spectra from the WFC3 Exposure Time Calculator (ETC) with temperatures ranging from 4060$-$9230 K. The closest matching model spectrum to HAT-P-32A ($T_{eff}$ = 6000 K) was the 5860 K model. We used this process to determine shifts along the dispersion axis over the course of the observations.       

\subsection{Spitzer/IRAC}

We obtained two transit observations of HAT-P-32Ab on UT 2012 November 18 and UT 2013 March 19 with the \textit{Spitzer} Infrared Array Camera (IRAC) 3.6 $\mu$m and 4.5 $\mu$m channels, respectively (\citealt{Werner04,Fazio04}). 
Each IRAC exposure was taken over integration times of 2 seconds in the 32 x 32 pixel subarray mode. 
We reduced the 3.6 and 4.5~$\mu$m \textit{Spitzer}/IRAC data using a custom data analysis pipeline which implements pixel-level decorrelation (PLD; \citealt{Deming15}), described fully in Baxter et al. (in prep). In summary, the pipeline performs a full search of the data reduction parameter space in order to determine the optimum aperture photometry, background subtraction, and centroiding.  The resulting photometric light curve is normalized to the out-of-transit flux, and errors are scaled with the photon noise. We clipped outliers with a sliding 4$\sigma$ median filter.

\subsection{Photometric Activity Monitoring}
\label{sec:activity}


\begin{deluxetable}{cccc}
\tabletypesize{\scriptsize}
\tablewidth{0pt}
\tablecolumns{4}
\tablecaption{Summary of photometric observations for HAT-P-32Ab\label{tab:ait}}

\tablehead{\colhead{Season} & \colhead{$N_{obs}$} & \colhead{Date Range} & \colhead{Sigma}  \\ \colhead{} & \colhead{} & \colhead{(HJD - 2,450,000)} & \colhead{(mag)}  }
\startdata 
 2014-15  &  79 & 56943--57114 & 0.00269  \\
 2015-16  &  82 & 57293--57472 & 0.00280  \\
 2016-17  &  55 & 57706--57843 & 0.00270  \\
 2017-18  &  13 & 58172--58288 & 0.00264  \\
 2018-19  &  41 & 58384--58510 & 0.00249  \\
\enddata
\end{deluxetable}
\label{tab:ait}

Stellar activity can mimic planetary signals and imprint spectral slopes and spurious absorption features in transmission spectra (e.g., \citealt{Pont13,McCullough14}). 
To assess whether stellar activity might impact the transit observations, we inspected available ground-based photometry from the All-Sky Automated Survey for Supernovae (ASAS-SN) (\citealt{Shappee14,Kochanek17,Rackham17}) and the Tennessee State University (TSU) Celestron 14-inch (C14) automated imaging telescope (AIT) at Fairborn Observatory. Since the ASAS-SN data set exhibits large scatter ($\sigma$ $\sim$ 10 mmag) and is dominated by noise, we only use the AIT observations in our analysis of the host star's activity levels. 

We acquired a total of 270 nightly observations of HAT-P-32A over the past five observing seasons from 2014-15 to 2018-19 (see e.g., \citealt{Henry99,Eaton03}). The first three observing seasons were discussed in \citet{Nikolov18}, where we provide details about the observing and data reduction procedures.
On the basis of those three observing seasons, we concluded that HAT-P-32A is constant on
night-to-night timescales within the precision ($\sim$2 mmag) of our observations and likely to be constant on year-to-year timescales.

The SBIG STL-1001E CCD camera on the AIT suffered a failure early in the 2017-18 observing
season and had to be replaced, resulting in an abbreviated fourth observing season. The
camera was replaced with another SBIG STL-1001E CCD to minimize instrumental shifts in
the long-term data. Nonetheless, we found that the 2017-18 and 2018-19 observing seasons
had seasonal mean differential magnitudes several milli-magnitudes different from the earlier
data. The observations are summarized in Table \ref{tab:ait}, but we have not included measurements
of the seasonal mean magnitudes because of the calibration uncertainties. We note
that the small nightly scatter in the new data is consistent with the star remaining
constant within the precision of our data on night-to-night timescales.

\begin{figure}
    \centering
    \includegraphics[scale=0.78]{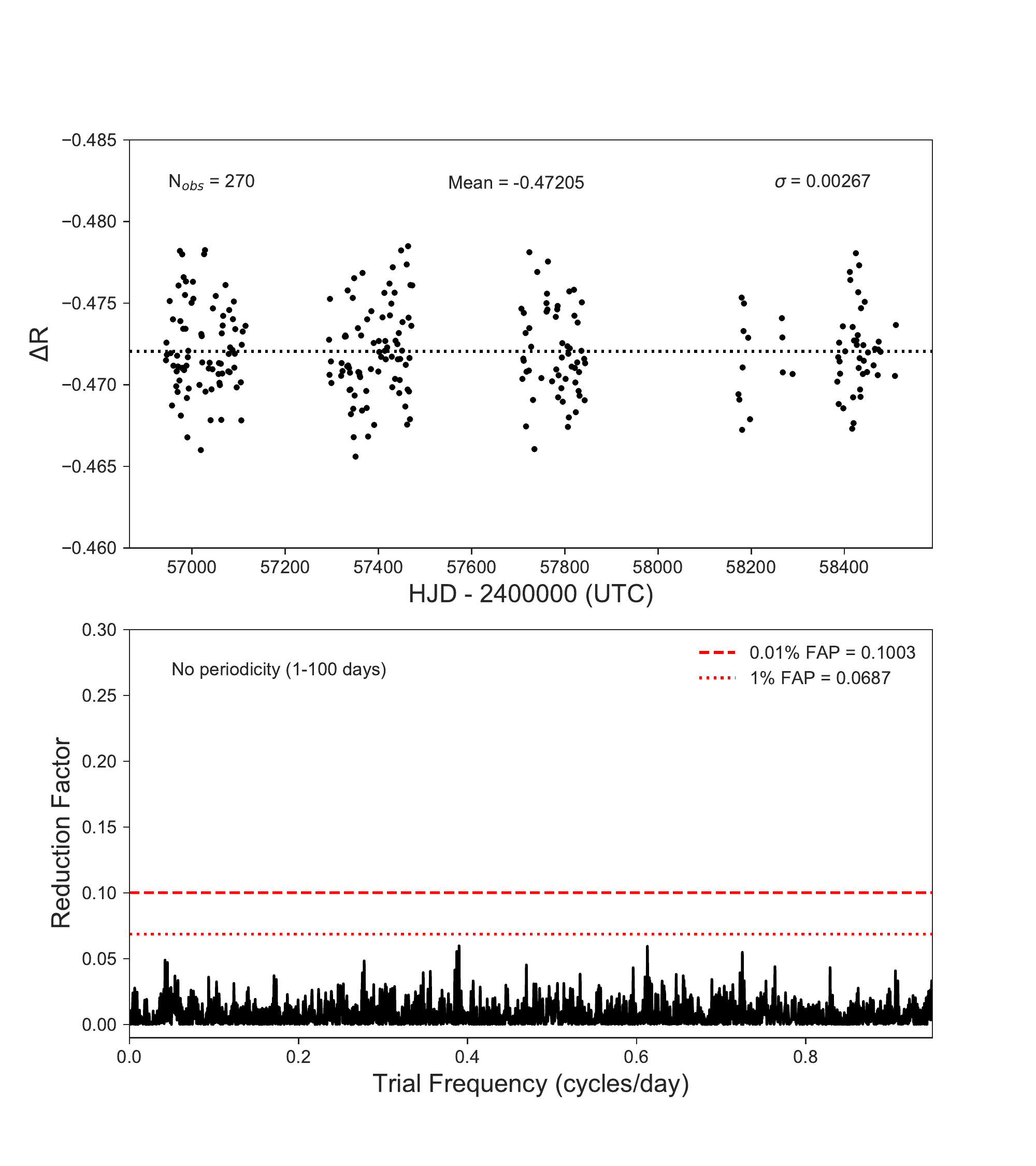}
        \caption{\textit{Top}: Photometry of HAT-P-32A across five observing seasons from 2014-15 to 2018-19, acquired in the Cousins R band with the TSU Celestron-14 AIT at Fairborn Observatory. The observations have been normalized so that all observing seasons have the same mean as the first season. \textit{Bottom}: Periodogram of the normalized 2014-2019 observations showing the lack of any significant periodicity between 1 and 100 days. We are therefore unable to detect any rotational variability in our observations.}
    \label{fig:ait}
\end{figure}

The complete HAT-P-32A AIT data set is plotted in the top panel of Figure \ref{fig:ait}, where the
data have been normalized so that each seasonal-mean differential magnitude is the same as the first observing season. The bottom panel shows a Lomb-Scargle periodogram (\citealt{Lomb76,Scargle82}) of our complete data set, which shows no evidence for any coherent periodicity between 1 and 100 days.

We further consider XMM-Newton observations taken on UT 2019 August 30 (P.I.: Sanz-Forcada). These observations reveal an X-ray flux of $L_{\rm X}=2\times 10^{29}$~erg\,s$^{-1}$ in the EPIC cameras using $d=291.5$~pc (Gaia DR2), in addition to the presence of two small flares (see further details in Sanz-Forcada et al., in prep.). EPIC cannot separate the A and B components of the HAT-P-32 system; so although the emission most likely comes from the A component, part of it might originate from the M dwarf companion. Considering this possibility, we checked observations from the optical monitor (OM) onboard XMM-Newton with the UVW2 filter ($\lambda = 1870-2370~\AA$). These observations indicate a low-level of activity in HAT-P-32A while the companion is not detected, reinforcing the idea that most of the X-ray emission originates from the A component of the system.
The UV and X-ray observations, which are most sensitive to the star's chromosphere, reveal some level of activity, while HAT-P-32A's photosphere (probed by the optical ground-based monitoring) appears quiet. Given these discrepant results, we decided to fit for activity in our retrievals as described in more detail in \S \ref{sec:retrieval}.

\section{\textit{HST} \& \textit{Spitzer} Light Curve Fits} 
\label{sec:analysis}

We extracted the 0.3$-$0.5 $\mu$m transmission spectrum of HAT-P-32Ab following the methods of \citealt{Sing11,Sing13}, \citealt{Nikolov14}, and \citealt{Alam18}. For each light curve, we simultaneously fit for the transit and systematic effects by fitting a two-component function consisting of a transit model multiplied by a systematics detrending model. The fitting procedure for the STIS, WFC3, and IRAC white light curves is described in \S \ref{sec:wlc}. The fitting procedure for the \textit{HST} spectroscopic light curves is detailed in \S \ref{sec:spec_lc}.  

\subsection{White Light Curves}
\label{sec:wlc}

\begin{figure}
    \centering
    \includegraphics[scale=0.78]{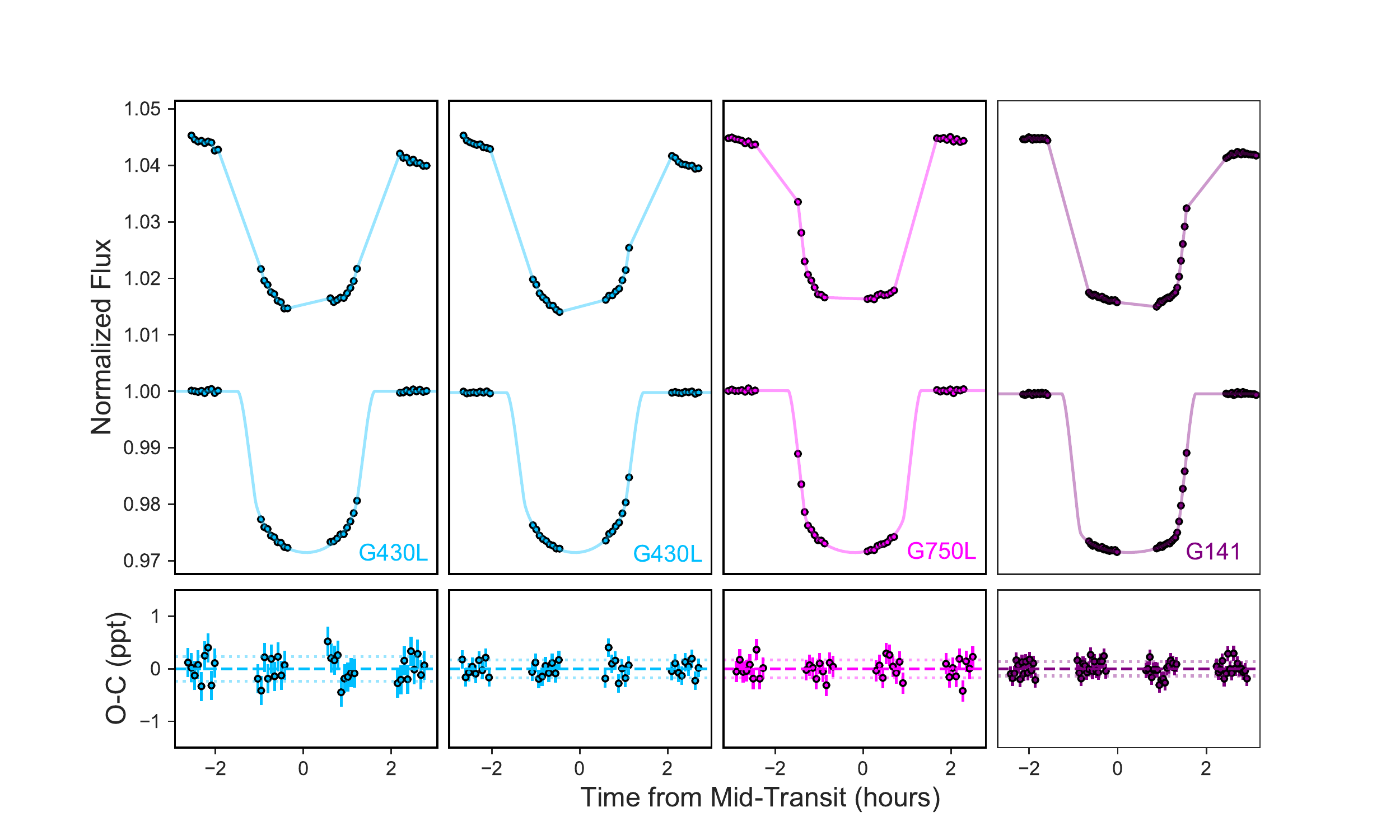}
    \caption{\textit{Top}: The raw and detrended white light curves (excluding the first orbit and the first exposure of each subsequent orbit) for each \textit{HST} visit in the STIS G430L (blue), STIS G750L (pink), and WFC3 G141 (purple) grisms. The best-fit analytical light curve model is overplotted. \textit{Bottom}: Transit fit residuals (in parts per thousand) with error bars.}
    \label{fig:wlc}
\end{figure}

\begin{figure}
    \centering
    \includegraphics[scale=0.37]{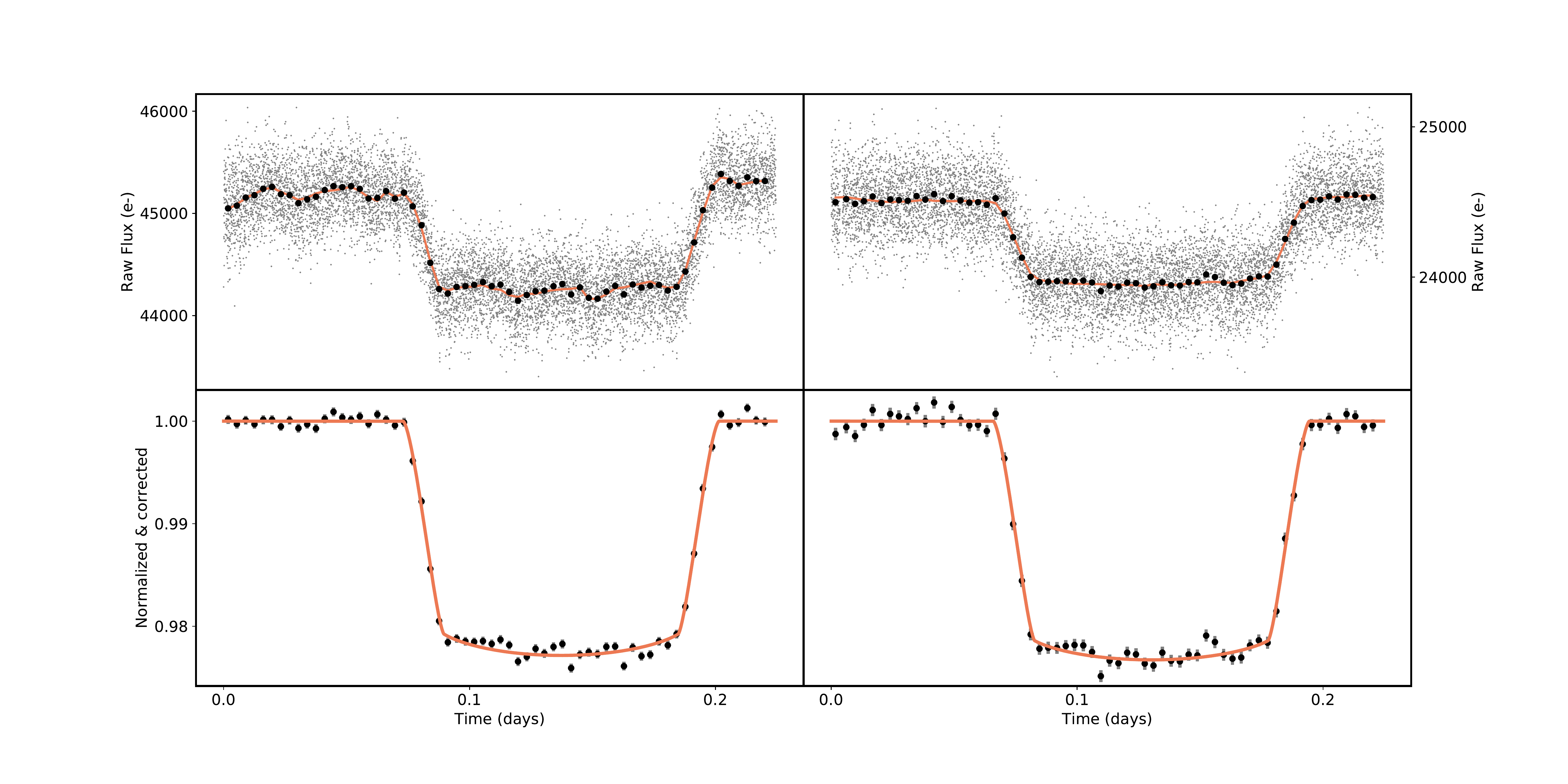}
    \caption{\textit{Top}: Raw flux (gray points) for the 3.6 $\mu$m (left) and 4.5 $\mu$m (right) \textit{Spitzer}/IRAC transit light curves, overlaid with the light curve binned to five minutes (black points).  \textit{Bottom}: Detrended light curves (black points) with the best-fit transit model (red line) overplotted.}
    \label{fig:spitzer}
\end{figure}

We produced the white light curves for the \textit{HST} and \textit{Spitzer} data sets by summing the flux of the stellar spectra across the full spectrum. We fit the white light curves using a complete analytic transit model \citep{Mandel02} parametrized by the mid-transit time $T_{0}$, orbital period $P$, inclination $i$, normalized planet semi-major axis $a/R_{\star}$, and planet-to-star radius ratio $R_{p}/R_{\star}$ (see \S \ref{sec:stis_lc} and \S \ref{sec:wfc3_lc} below). The raw and detrended white light curves are shown in Figures \ref{fig:wlc} and \ref{fig:spitzer}. The derived system parameters for HAT-P-32Ab from these fits are given in Table \ref{tab:params}.  


\begin{deluxetable*}{cccccc}
\tabletypesize{\scriptsize}
\tablewidth{0pt}
\tablecolumns{6}
\tablecaption{White light curve derived system parameters for HAT-P-32Ab\label{tab:params}}
\tablehead{\colhead{}  & \colhead{STIS G430L (visit 72)}  & \colhead{STIS G430L (visit 73)}  & \colhead{STIS 750L (visit 74)} & \colhead{WFC3 G141} & \colhead{\textit{Spitzer}/IRAC\tablenotemark{a}}}
\startdata 
Period, $P$ [days]                     &   2.15 (fixed)          		&  2.15 (fixed)           &  2.15 (fixed)      		&  2.15 (fixed)  & 2.15 (fixed) 	       \\
Orbital inclination, $i$ [$^{\circ}$]  &   89.53 $\pm$ 1.02               &  88.97 $\pm$ 0.20   	  &  88.50 $\pm$ 1.02         &  87.78 $\pm$ 0.5   &   89.55 $\pm$ 0.5    \\
Orbital eccentricity, $e$              &   0.0 (fixed)                  &  0.0 (fixed)     	      &  0.0 (fixed)      		&  0.0 (fixed)     &  0.0 (fixed)             \\
Scaled semi-major axis, $a/R_{\star}$  &   5.98 $\pm$ 0.05          	&  5.96 $\pm$ 0.06       &  6.22 $\pm$ 0.11       &  6.17 $\pm$ 0.03   &  6.13 $\pm$ 0.04   \\
Radius ratio, $R_{p}/R_{\star}$ 	   &   0.1516 $\pm$ 0.0002          &  0.1510 $\pm$ 0.0002    &  0.1499 $\pm$ 0.0003    &  0.1511 $\pm$ 0.0002  & 0.1502 $\pm$ 0.0009 \\
\enddata           
\tablenotetext{a}{The values reported in this column are the weighted mean of the fitted parameters from the \textit{Spitzer} 3.6 $\mu$m and 4.5 $\mu$m observations. The reported $R_{p}/R_{\star}$ values are weighted mean of the radius ratio corrected for dilution from the companion to HAT-P-32A, as described in \S \ref{sec:irac_lc}. }
\end{deluxetable*}

\label{tab:params}

\subsubsection{STIS}
\label{sec:stis_lc}

To produce the STIS white light curves, we summed each spectrum over the complete bandpasses (2892$-$5700 \AA~ for the G430L grating; 5240$-$10270 \AA~ for the G750L grating) and derived photometric uncertainties  based on pure photon statistics. The raw white light curves exhibited typical STIS systematic trends related to the spacecraft's orbital motion (\citealt{Gilliland99,Brown01}). We detrended these instrumental systematics by applying orbit-to-orbit flux corrections that account for the spacecraft orbital phase ($\phi_{t}$), drift of the spectra on the detector ($x$ and $y$), the shift of the stellar spectrum cross-correlated with the first spectrum of the time series ($\omega$), and time ($t$). Following common practice, we excluded the first orbit and the first exposure of each subsequent orbit because these data were taken while the telescope was thermally relaxing into its new pointing position and have unique, complex systematics \citep{Huitson13}. 

We then generated a family of systematics models spanning all possible combinations of detrending variables and performed separate fits including each systematics model in the two-component function. We assumed zero eccentricity, fixed $P$ to the value given in \citet{Hartman11}, and fit for $i$, $a/R_{\star}$, $T_{0}$, $R_{p}/R_{\star}$, instrument systematic trends, and stellar baseline flux. We derived the four non-linear stellar limb darkening coefficients based on 3D stellar models \citep{Magic15} and adopted these values as fixed parameters in the transit fits. 
We used a Levenberg-Marquardt least-squares fitting routine \citep{Markwardt09} to determine the best-fit parameters of the combined transit+systematics function. We marginalized over the entire set of functions following the \citet{Gibson14} framework, and selected which systematics model to use based on the lowest Akaike Information Criterion (AIC; \citealt{Akaike74}) value \citep{Nikolov14}. See Appendix \ref{sec.sys} for further details.

\subsubsection{WFC3}
\label{sec:wfc3_lc}

To produce the WFC3 white light curve, we integrated the flux in each spectrum over the full G141 grism bandpass (1.1$-$1.7 $\mu$m). The raw WFC3 white light curves exhibited typical instrumental systematic trends associated with a visit-long linear slope and the known ``ramping" effect in which the flux asymptotically increases over each orbit due to residual charge on the detector from previous exposures (\citealt{Deming13,Huitson13,Zhou17}). In accordance with common practice, the first orbit and the first exposure of each subsequent orbit were excluded due to the well-known charge-trapping ramp systematics for WFC3 (e.g., \citealt{Kreidberg15b,Evans17}). 

We then fit the light curve with an analytical model that takes into account the ramping effect and the thermal breathing of \textit{HST}. We fixed $e$ to zero and $P$ to the value from \citealt{Hartman11}, and fit for $i$, $a/R_{\star}$, $R_{p}/R_{\star}$, $T_{0}$, and instrument systematics. We derived the theoretical limb darkening coefficients based on the 3D stellar models of \citet{Magic15}. As in our analysis of the STIS light curves (see \S \ref{sec:stis_lc}), we generated a family of systematics models, detrended the raw WFC3 light curve by performing separate fits to each model, and marginalized over the entire set of functions (c.f. \citealt{Wakeford16} for further details). We used the lowest AIC value to select which model to use. For further details on the systematics model selection, see Appendix \ref{sec.sys}. 

\subsubsection{IRAC}
\label{sec:irac_lc}

We fit the cleaned and normalized IRAC light curves with a {\tt batman} transit model \citep{Kreidberg15} in combination with the PLD systematic model and temporal ramp, resulting in 14 free parameters (four {\tt batman}, nine PLD, and one temporal ramp). Furthermore, we fixed the eccentricity $e$ to zero and the orbital period $P$ to the literature value of 2.15 days \citep{Hartman11}, and fit for $i$, $a/R_{\star}$, $T_{0}$, and $R_{p}/R_{star}$. We used the linear limb darkening law to calculate the theoretical limb darkening coefficients using the 1D ATLAS code presented in \citet{Sing10}. Posteriors for all 14 free parameters were calculated using the Markov Chain Monte Carlo (MCMC) script {\tt emcee} \citep{Foreman-Mackey13}. The final transit parameters presented in Table \ref{tab:params} are the result of a second MCMC, where the semi-major axis $a/R_{\star}$ and the inclination $i$ were varied within Gaussian priors from the median and standard deviation of the initial fits. 

From these fits, we derive $R_{p}/R_{\star}$ values of 0.14663 $\pm$ 0.00034 and 0.14866 $\pm$ 0.00067 for the 3.6 $\mu$m and 4.5 $\mu$m IRAC channels, respectively. Considering the 1.2" x 1.2" pixel size for the \textit{Spitzer} 32x32 subarray images, we must correct for dilution from the M dwarf companion to HAT-P-32A.  We applied the dilution correction derived in \citet{Stevenson14}: 

\begin{equation}
    \delta_{true}(\lambda) = \delta_{obs}(\lambda)[1 + g(\beta,\lambda)\frac{F_{B}}{F_{A}}]
\end{equation}

\noindent where $\delta_{true}(\lambda)$ is the true (undiluted) transit depth, $\delta_{obs}(\lambda)$ is the observed (diluted) transit depth, $g(\beta,\lambda)$ is wavelength-dependent companion flux fraction inside a photometric aperture of size $\beta$, $F_{B}$ is the flux of the companion star, and $F_{A}$ is the in-transit flux of the primary star. To account for the third light contribution in the \textit{Spitzer} images, we use the dilution factors of $(F_{B}/F_{A})_{3.6}$ = 0.050$\pm$0.020 and $(F_{B}/F_{A})_{4.5}$ = 0.053$\pm$0.020 from \citet{Zhao14} and estimate $g(\beta,\lambda)$ for an aperture radius of 2.5 pixels using the IRAC point response function (PRF)\footnote{\url{https://irsa.ipac.caltech.edu/data/SPITZER/docs/irac/calibrationfiles/psfprf/}} at 1/5th pixel sampling. The resulting $R_{p}/R_{\star}$ values corrected for dilution are reported in Tables \ref{tab:params} and \ref{tab:tr_spec}.

\subsection{Spectroscopic Light Curves}
\label{sec:spec_lc}


\begin{deluxetable*}{cccccc}
\tabletypesize{\scriptsize}
\tablewidth{0pt}
\tablecolumns{6}
\tablecaption{Broadband \textit{HST}+\textit{Spitzer} transmission spectrum for HAT-P-32Ab \& adopted non-linear (\textit{HST}) and linear (\textit{Spitzer}) limb darkening coefficients 
\label{tab:tr_spec}}

\tablehead{\colhead{$\lambda$ (\AA)}  & \colhead{$R_{p}/R_{*}$} & \colhead{$c_{1}$}  & \colhead{$c_{2}$}  & \colhead{$c_{3}$}  & \colhead{$c_{4}$} }

\startdata 
2900$-$3300   & 0.15466 $\pm$ 0.00158 & 0.3152 & 0.4420  & 0.4813   & -0.3167  \\ 
3300$-$3700   & 0.15281 $\pm$ 0.00088 & 0.4052 & 0.6943  & -0.2319  &  0.0273  \\                                      
3700$-$3950   & 0.15203 $\pm$ 0.00073 & 0.4069 & 0.5814  & 0.0073   & -0.1117  \\                                      
3950$-$4200   & 0.15225 $\pm$ 0.00054 & 0.3991 & 0.5794  & 0.0046   & -0.0954  \\                                      
4200$-$4350   & 0.15084 $\pm$ 0.00093 & 0.4025 & 0.5039  & 0.0782   & -0.1137  \\                                      
4350$-$4500   & 0.15104 $\pm$ 0.00068 & 0.4998 & 0.3418  & 0.1836   & -0.1546  \\                                      
4500$-$4650   & 0.15126 $\pm$ 0.00066 & 0.5702 & 0.2601  & 0.0992   & -0.0640  \\                                      
4650$-$4800   & 0.15104 $\pm$ 0.00063 & 0.5660 & 0.3170  & -0.0081  &  -0.0204 \\                                      
4800$-$4950   & 0.15083 $\pm$ 0.00065 & 0.6888 & 0.1103  & 0.1042   & -0.0767  \\                                      
4950$-$5100   & 0.15093 $\pm$ 0.00049 & 0.6243 & 0.1792  & 0.0510   & -0.0290  \\                                      
5100$-$5250   & 0.15137 $\pm$ 0.00059 & 0.6077 & 0.1870  & 0.0812   & -0.0633  \\                                      
5250$-$5400   & 0.15183 $\pm$ 0.00049 & 0.6782 & 0.0034  & 0.2548   & -0.1367  \\                                      
5400$-$5550   & 0.15080 $\pm$ 0.00051 & 0.7363 & -0.0980 & 0.2614   & -0.1063  \\                                      
5550$-$5700   & 0.15128 $\pm$ 0.00060 & 0.7356 & -0.1217 & 0.2683   & -0.1016  \\                                      
5700$-$6000   & 0.15077 $\pm$ 0.00070 & 0.7728 & -0.2053 & 0.3104   & -0.1130  \\ 
6000$-$6300   & 0.15105 $\pm$ 0.00058 & 0.7964 & -0.2947 & 0.3789   & -0.1381  \\                                      
6300$-$6500   & 0.15057 $\pm$ 0.00122 & 0.8037 & -0.3285 & 0.4036   & -0.1533  \\                                      
6500$-$6700   & 0.14924 $\pm$ 0.00075 & 0.8718 & -0.4706 & 0.4820   & -0.1819  \\                                      
6700$-$6900   & 0.14933 $\pm$ 0.00072 & 0.8333 & -0.4336 & 0.4641   & -0.1631  \\                                      
6900$-$7100   & 0.15066 $\pm$ 0.00069 & 0.8462 & -0.4889 & 0.5201   & -0.1886  \\                                      
7100$-$7300   & 0.15121 $\pm$ 0.00097 & 0.8461 & -0.4985 & 0.5090   & -0.1780  \\                                      
7300$-$7500   & 0.15022 $\pm$ 0.00058 & 0.8321 & -0.4776 & 0.4849   & -0.1740  \\                                      
7500$-$7700   & 0.15084 $\pm$ 0.00071 & 0.8520 & -0.5558 & 0.5665   & -0.2086  \\                                      
7700$-$8100   & 0.14905 $\pm$ 0.00073 & 0.8573 & -0.5815 & 0.5666   & -0.2010  \\                                      
8100$-$8350   & 0.15021 $\pm$ 0.00110 & 0.8645 & -0.6135 & 0.5794   & -0.2024  \\                                      
8350$-$8600   & 0.15080 $\pm$ 0.00122 & 0.8574 & -0.6348 & 0.6070   & -0.2167  \\                                      
8600$-$8850   & 0.15013 $\pm$ 0.00110 & 0.8560 & -0.6383 & 0.5907   & -0.2071  \\                                      
8850$-$9100   & 0.15105 $\pm$ 0.00189 & 0.8622 & -0.6681 & 0.6155   & -0.2188  \\                                      
9100$-$9500   & 0.14906 $\pm$ 0.00146 & 0.8598 & -0.6768 & 0.6389   & -0.2305  \\                                      
9500$-$10200  & 0.14939 $\pm$ 0.00113 & 0.8479 & -0.6659 & 0.6118   & -0.2182  \\                                      
11190$-$11470 & 0.15071 $\pm$ 0.00035 & 0.6341 & -0.2157 & 0.1764   & -0.0625 \\  
11470$-$11750 & 0.15068 $\pm$ 0.00031 & 0.6336 & -0.2103 & 0.1587   & -0.0553 \\  
11750$-$12020 & 0.15136 $\pm$ 0.00033 & 0.6311 & -0.2011 & 0.1333   & -0.0413 \\  
12020$-$12300 & 0.15119 $\pm$ 0.00030 & 0.6282 & -0.1673 & 0.0809   & -0.0204 \\  
12300$-$12580 & 0.15055 $\pm$ 0.00028 & 0.6318 & -0.1698 & 0.0748   & -0.0191 \\  
12580$-$12860 & 0.15065 $\pm$ 0.00032 & 0.6566 & -0.1844 & 0.0366   & -0.0005 \\  
12860$-$13140 & 0.15048 $\pm$ 0.00035 & 0.6480 & -0.1651 & 0.0284   &  0.0051 \\  
13140$-$13420 & 0.15148 $\pm$ 0.00027 & 0.6588 & -0.1768 & 0.0249   &  0.0089 \\  
13420$-$13700 & 0.15204 $\pm$ 0.00033 & 0.6724 & -0.1969 & 0.0252   &  0.0125 \\  
13700$-$13980 & 0.15168 $\pm$ 0.00030 & 0.6987 & -0.2291 & 0.0299   &  0.0157 \\  
13980$-$14260 & 0.15182 $\pm$ 0.00030 & 0.7189 & -0.2589 & 0.0426   &  0.0140 \\  
14260$-$14540 & 0.15202 $\pm$ 0.00029 & 0.7400 & -0.3024 & 0.0668   &  0.0091 \\  
14540$-$14820 & 0.15122 $\pm$ 0.00039 & 0.7750 & -0.3619 & 0.1059   & -0.0025 \\  
14820$-$15090 & 0.15180 $\pm$ 0.00034 & 0.8033 & -0.4316 & 0.1561   & -0.0152 \\  
15090$-$15370 & 0.15067 $\pm$ 0.00036 & 0.8629 & -0.5486 & 0.2411   & -0.0365 \\  
15370$-$15650 & 0.15172 $\pm$ 0.00039 & 0.8773 & -0.6057 & 0.3004   & -0.0586 \\  
15650$-$15930 & 0.15114 $\pm$ 0.00036 & 0.8491 & -0.5982 & 0.3194   & -0.0704 \\  
15930$-$16210 & 0.15015 $\pm$ 0.00039 & 0.9445 & -0.8091 & 0.5039   & -0.1343 \\  
16210$-$16490 & 0.14947 $\pm$ 0.00042 & 0.9501 & -0.8296 & 0.5057   & -0.1253 \\  
36000 		  & 0.14820 $\pm$ 0.00078 & 0.1816 $\pm$ 0.0048 & -- & -- & -- \\  
45000 		  & 0.15020 $\pm$ 0.00087 & 0.1614 $\pm$ 0.0051 & -- & -- & -- \\
\enddata
\end{deluxetable*}
\label{tab:tr_spec}

\begin{figure}
    \centering
    \includegraphics[scale=0.80]{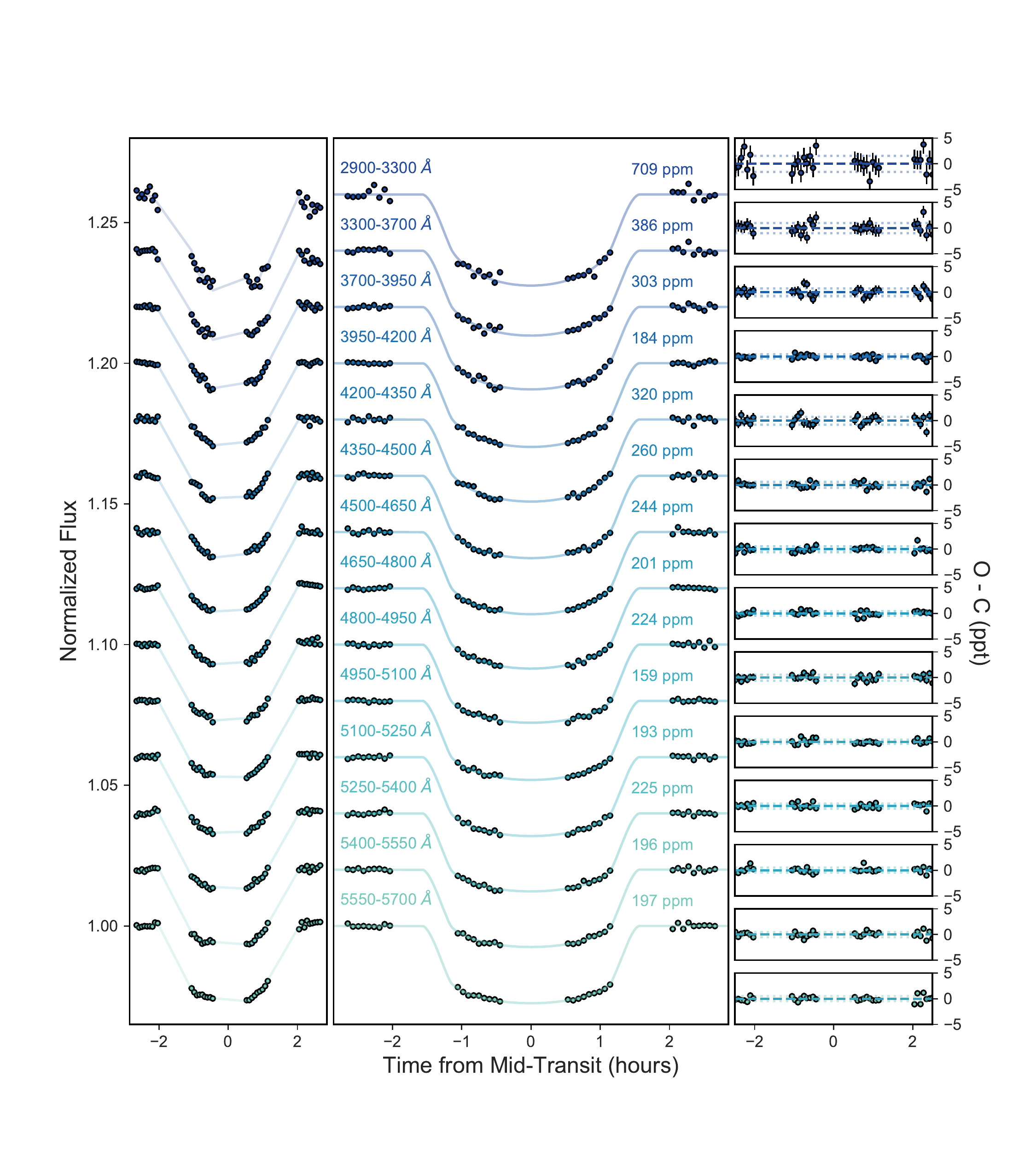}
    \caption{\textit{HST}/STIS G430L (visit 72) spectrophotometric light curves. The common mode corrected raw (left) and detrended (middle) light curves shown for each wavelength bin are offset vertically by an arbitrary constant for clarity. The observed minus computed residuals (parts per thousand) with error bars are shown in the right panel.}
    \label{fig:v72_bins}
\end{figure}

\begin{figure}
    \centering
    \includegraphics[scale=0.80]{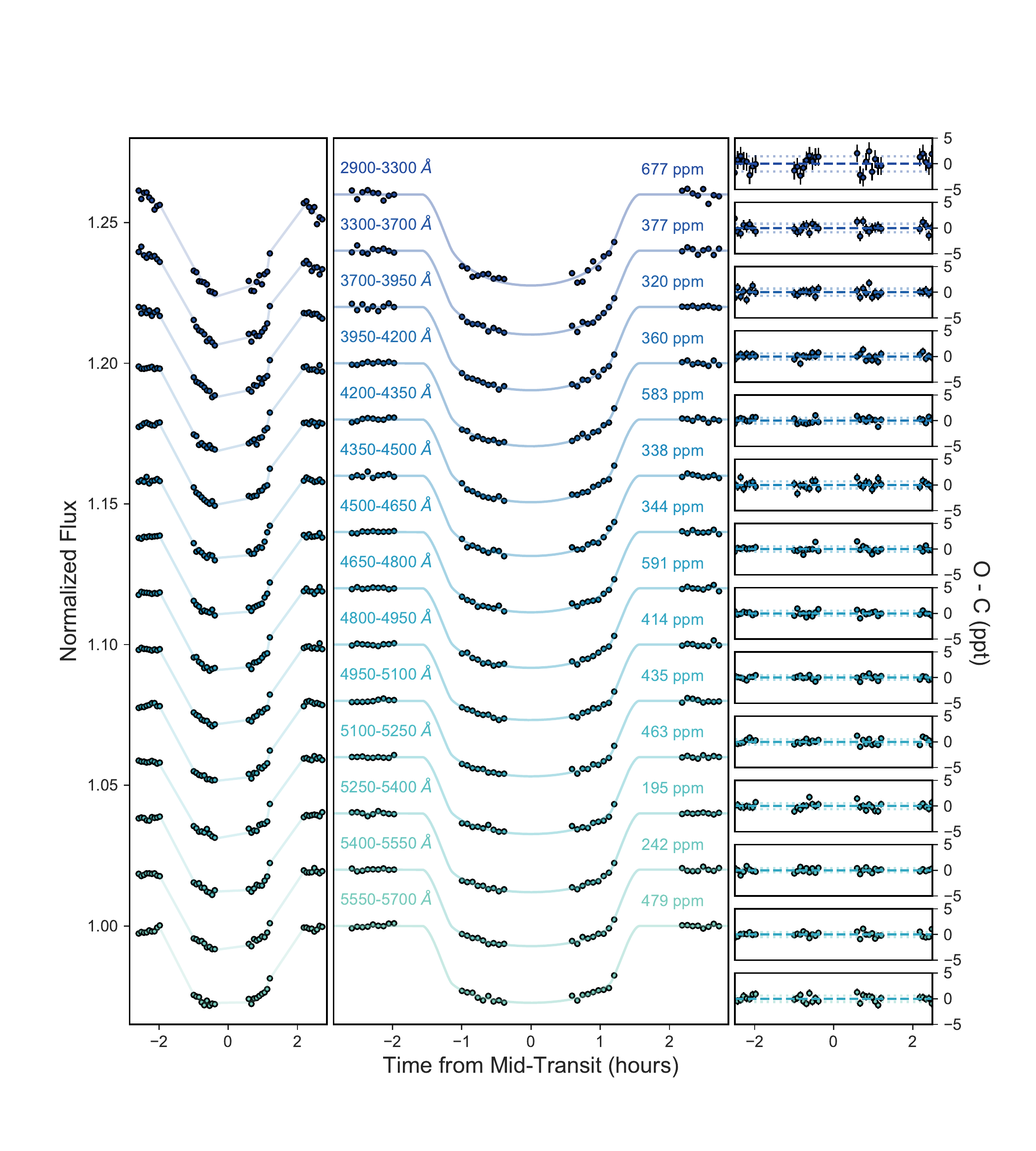}
    \caption{Same as Figure \ref{fig:v72_bins}, but for \textit{HST}/STIS G430L visit 73.}
    \label{fig:v73_bins}
\end{figure}

\begin{figure}
    \centering
    \includegraphics[scale=0.75]{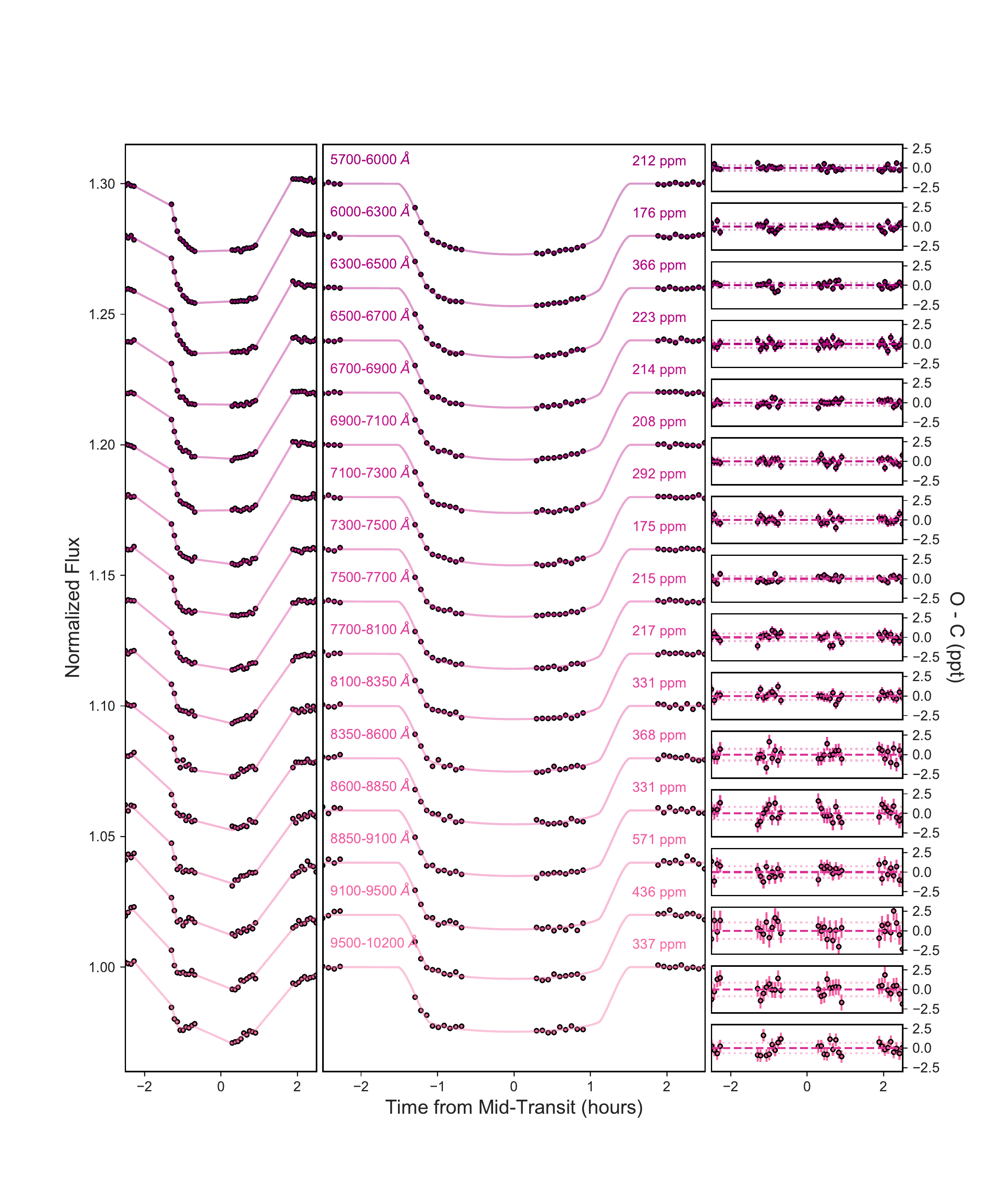}
    \caption{Same as Figure \ref{fig:v72_bins}, but for \textit{HST}/STIS G750L visit 74.}
    \label{fig:v74_bins}
\end{figure}

\begin{figure}
    \centering
    \includegraphics[scale=0.70]{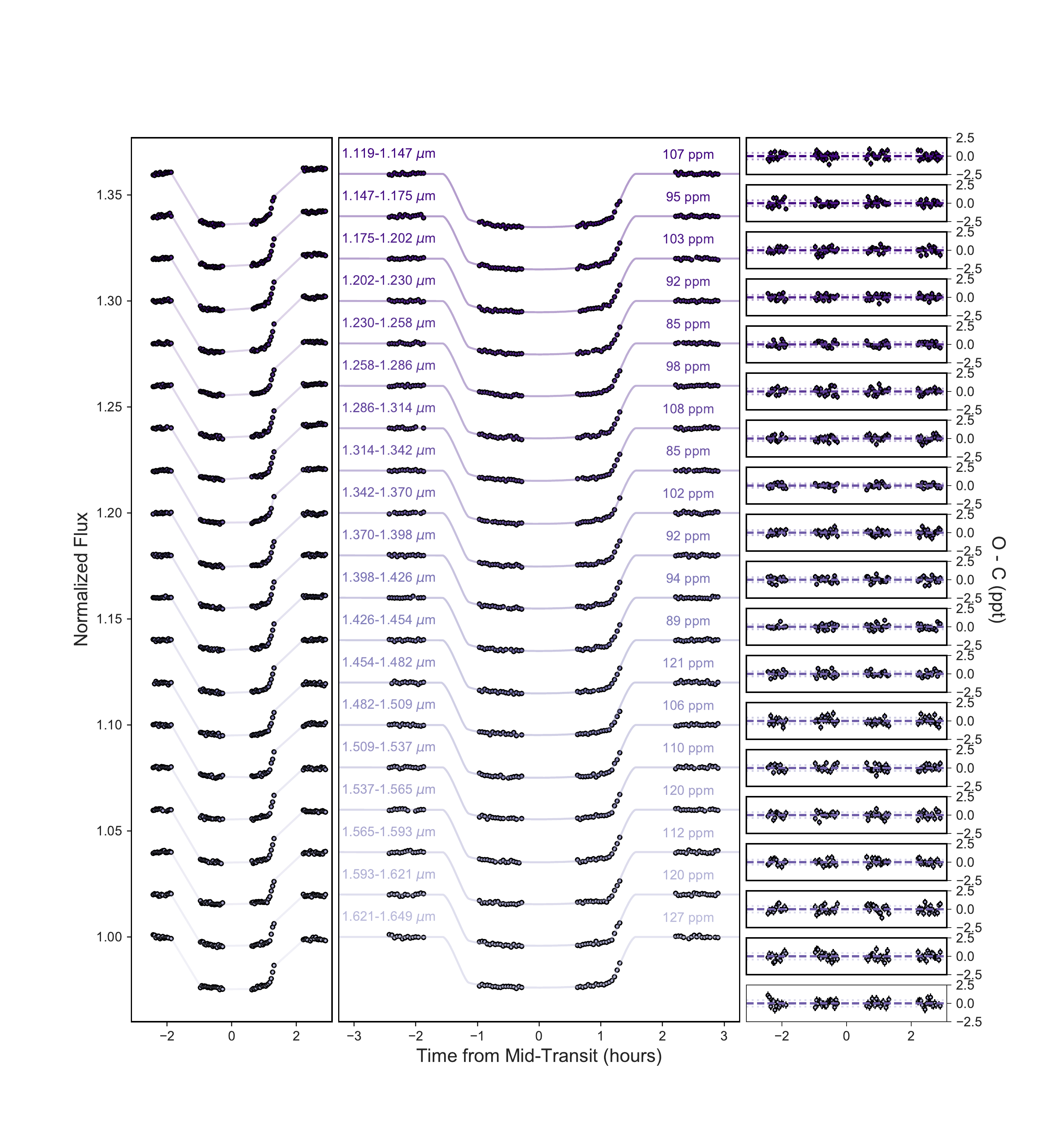}
    \caption{Same as Figure \ref{fig:v72_bins}, but for \textit{HST}/WFC3 G141 visit 01.}
    \label{fig:v01_bins}
\end{figure}

To produce the spectroscopic light curves, we binned the STIS and WFC3 spectra into 49 spectrophotometric channels between 0.3$-$1.7 $\mu$m. The resulting binned light curves are shown in Figures \ref{fig:v72_bins}, \ref{fig:v73_bins}, \ref{fig:v74_bins}, and \ref{fig:v01_bins}. We produced 30 STIS spectrophotometric light curves by summing the flux of the stellar spectra in bins with widths ranging from 0.015 to 0.04 $\mu$m. We used a range of bin widths to achieve similar fluxes in each spectroscopic channel as well as avoid stellar absorption lines. To generate the 19 WFC3 spectroscopic light curves, we summed the flux of the stellar spectra in uniformly sized bins of six pixels (0.028 $\mu$m) each. 

We performed a common mode correction to remove wavelength-independent systematic trends from each spectroscopic channel and reduce the amplitude of the observed \textit{HST} breathing systematics. Common mode trends are computed by dividing the raw flux of the white light curve in each grating by the best-fitting transit model. We applied the common mode correction by dividing each spectrophotometric light curve by the computed common mode flux, which may cause offsets between the independent data sets. We then fit each spectroscopic light curve following the same procedure as the white light curves (see \S \ref{sec:stis_lc} and \S \ref{sec:wfc3_lc} for details), but fixed $T_{0}$ to the white light curve best-fit value. We also fixed $i$ and $a/R_{\star}$ to the values from \citet{Hartman11} to reduce the effect of instrumental offsets between the  different datasets. The limb darkening coefficients were fixed to the computed theoretical values for each wavelength bin (see Table \ref{tab:tr_spec}). The measured $R_{p}/R_{\star}$ values for each spectroscopic channel are presented in Table \ref{tab:tr_spec}. 



\section{Results} \label{sec:results}

We construct the optical to infrared transmission spectrum for HAT-P-32Ab measured from 0.3$-$5 $\mu$m by combining the STIS, WFC3, and \textit{Spitzer} observations. 
The broadband spectrum (Table \ref{tab:tr_spec}) compared to previous atmospheric observations and forward models (\citealt{Goyal17,Goyal19}) is presented in Figure \ref{fig:tr_spec}. In this section, we characterize the shape and slope of the transmission spectrum compared to previous atmospheric observations (\S \ref{sec:tr_spec}) and present an interpretation of the planet's atmospheric structure and composition based on fits to a grid of 1D radiative-convective equilibrium models (\S \ref{sec:fits}) and retrievals (\S \ref{sec:retrieval}). 

\subsection{HST+Spitzer Transmission Spectrum \& Comparison with Previous Results}
\label{sec:tr_spec}

The optical to infrared transmission spectrum of HAT-P-32Ab is characterized by a weak H$_{2}$O absorption feature at 1.4 $\mu$m, no evidence of Na {\sc i} or K {\sc i} alkali absorption features, and a steep slope in the blue optical. This continuum slope may be due to the presence of an optical opacity source in the atmosphere of this planet, which \citet{Mallonn17} predict could be magnesium silicate aerosols. Additionally, we note that the reddest spectroscopic channels of the WFC3 observations ($\sim$1.57$-$1.65 $\mu$m) present a steep slope in the H$_{2}$O bandhead at $\sim$1.6 $\mu$m. This feature is also present in the independently reduced WFC3 results of \citet{Damiano17}, suggesting that it may be physical in nature and not an artifact of the data reduction process. This feature is not well modeled by the best-fitting ATMO models (\S \ref{sec:fits}) or PLATON retrievals (\S \ref{sec:retrieval}) 
and we note that it has been observed for other planets, such as the HAT-P-26b \citep{Wakeford17} and WASP-79b \citep{Sotzen19}.

There are several other measured transmission spectra for HAT-P-32Ab in addition to the \textit{HST} spectrum reported here, including observations from Gemini/GMOS \citep{Gibson13}, LBT/MODS \citep{Mallonn16a}, GTC/OSIRIS \citep{Nortmann16}, and LBC/LBT \citep{Mallonn17}.  Figure \ref{fig:tr_spec} shows our results compared to previously published optical and near-infrared transmission spectra. 
Cloud-free atmospheric models predict Na {\sc i} at 5893 \AA~and K {\sc i} at 7665 \AA, but ground-based optical transmission spectra of HAT-P-32Ab show no evidence of these pressure-broadened absorption features in addition to a Rayleigh-scattering slope (\citealt{Gibson13,Mallonn16a,Mallonn16b,Nortmann16,Tregloan17}). We varied the size of the spectroscopic channels centered on  Na {\sc i} and K {\sc i} to search for absorption signatures from these species and confirm no evidence of these features in the spectrum at the precision level of our data.

Our STIS, WFC3, and \textit{Spitzer} measurements are consistent with these previous ground-based observations in terms of the slope and shape of the transmission spectrum, as well as the  $R_{p}/R_{\star}$ baseline. Small offsets among data sets can be attributed to systematic errors, different data reduction techniques, and the challenges of measuring absolute transit depths from observations taken during different epochs as the stellar photosphere evolves (e.g., \citealt{Stevenson14,Kreidberg15b}). The agreement in the HAT-P-32Ab absolute transit depth measurements over several epochs, using ground-based as well as space-borne facilities, and with different instruments susceptible to different systematic effects reiterates the lack of variability in the photosphere of the stellar host (\S \ref{sec:activity}). 

 

\subsection{Fits to Forward Atmospheric Models}
\label{sec:fits}

\begin{figure}
    \centering
    \includegraphics[scale=0.75]{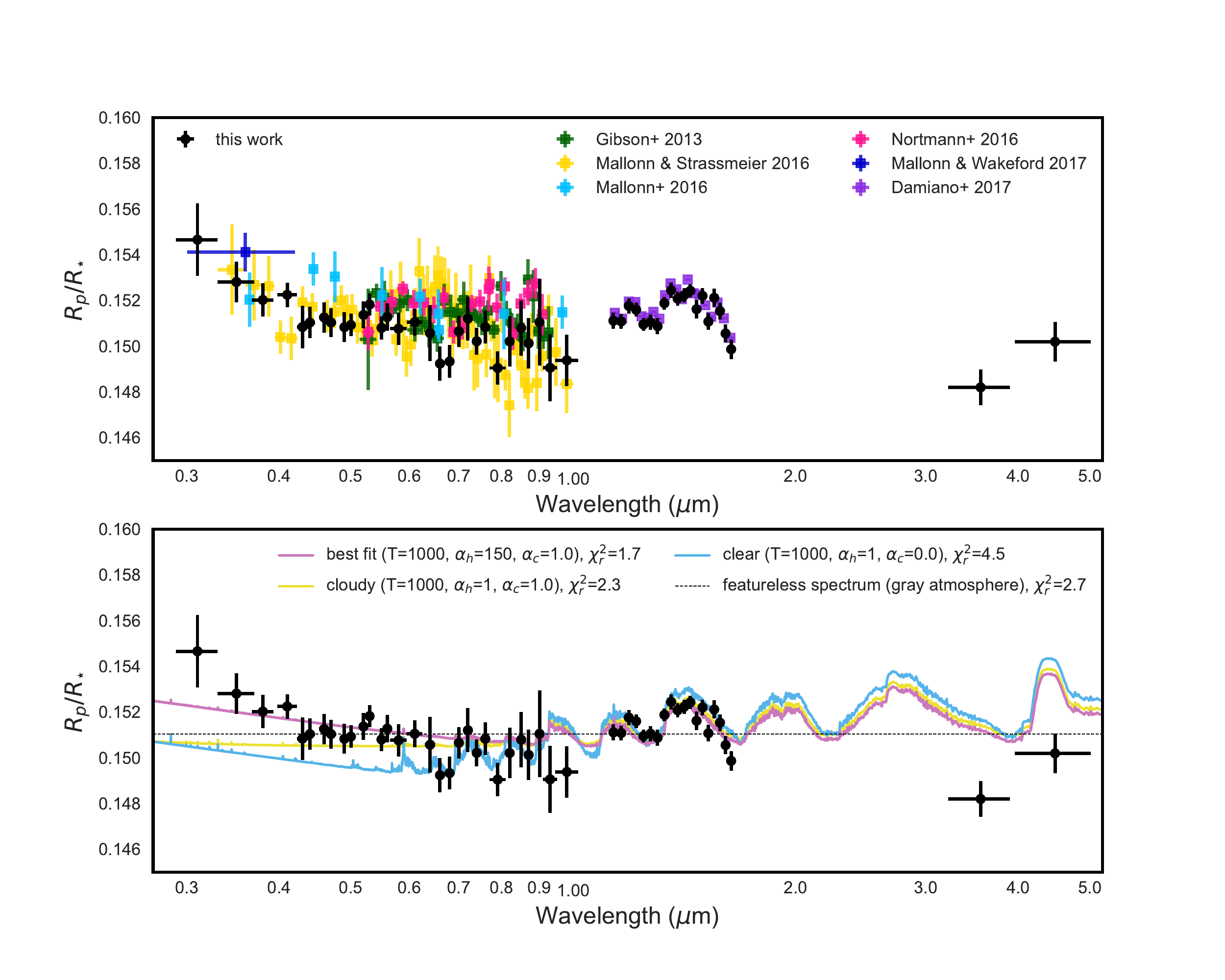}
    \caption{\textit{Top:} Broadband transmission spectrum for HAT-P-32Ab from \textit{HST} STIS+WFC3 and \textit{Spitzer} IRAC (black points). Ground-based measurements from \citealt{Gibson13} (green), \citealt{Mallonn16b} (yellow), \citealt{Mallonn16a} (light blue),
    \citealt{Nortmann16} (pink), \citealt{Mallonn17} (dark blue), and \citealt{Damiano17} (purple) are shown for comparison. \textit{Bottom}: A subset of the best-fitting theoretical models (lines) fit to the broadband transmission spectrum.  The increase in transit depth near 1.4 $\mu$m corresponds to a near-infrared H$_{2}$O bandhead. The average $R_{p}/R_{\star}$ baseline of the transmission spectrum (dashed black line) is shown for reference. }
    \label{fig:tr_spec}
\end{figure}

We compare our observed \textit{HST}+\textit{Spitzer} transmission spectrum (Figure \ref{fig:tr_spec}) to the publicly available generic grid of forward model transmission spectra presented in \citealt{Goyal17,Goyal19}. The 1D radiative-convective equilibrium models are produced using {\tt ATMO} (\citealt{Amundsen14,Tremblin15,Tremblin16,Drummond16}), computed assuming isothermal pressure-temperature ($P - T$) profiles and condensation without rainout (local condensation). The models include opacities due to H$_{2}$-H$_{2}$, H$_{2}$-He collision induced absorption, H$_{2}$O, CO$_{2}$, CO, CH$_{4}$, NH$_{3}$, Na, K, Li, Rb, Cs, TiO, VO, FeH, CrH, PH$_{3}$, HCN, C$_{2}$H$_{2}$, H$_{2}$S, and SO$_{2}$. The pressure broadening sources for these species are tabulated in \citet{Goyal17}. 

The entire generic {\tt ATMO} grid comprises 56,320 forward model transmission spectra for 22 equilibrium temperatures (400$-$2600 K in steps of 100 K), four planetary gravities (5, 10, 20, 50 m/s$^{2}$), five metallicities (1, 10, 50, 100, 200 x solar), and four C/O ratios (0.35, 0.56, 0.7, 1.0), as well as varying degrees of haziness (1, 10, 150, 1100) and cloudiness (0.0, 0.06, 0.20, 1.0). 
Gray scattering clouds are included in the models using the H$_{2}$ cross-section at 350 nm as a reference wavelength; the varying degrees of cloudiness are a multiplicative factor to this value. 

We fit the generic {\tt ATMO} model grid scaled to $g$ = 5 m/s$^{2}$ to the observed spectrum by computing the mean model prediction for the wavelength range of each spectroscopic channel (see Table \ref{tab:tr_spec}) and performing a least-squares fit of the band-averaged model to the spectrum. In the fitting procedure, we preserved the shape of the model by allowing the vertical offset in $R_{p}/R_{\star}$ between the spectrum and model to vary while holding all other parameters fixed. The number of degrees of freedom for each model is $n - m$, where $n$ is the number of data points and $m$ is the number of fitted parameters. Since $n$ = 51 and $m$ = 1, the number of degrees of freedom for each model is constant. From the fits, we quantified our model selection by computing the $\chi^{2}$ statistic.

The best-fitting model is shown in the bottom panel of Figure \ref{fig:tr_spec}, which also shows a flat model, and representative cloudy and clear atmosphere models for reference. The best fitting model ($\chi_{r}^{2}$ = 1.7) corresponds to a cloudy ($\alpha_{cloud}$ = 1.0) and slightly hazy ($\alpha_{haze}$ = 150) atmosphere, with a temperature of $T$ = 1000 K, super-solar metallicity ([Fe/H] = +1.7), and sub-solar C/O (C/O = 0.35). The selected clear ($\chi_{r}^{2}$ = 4.5) and cloudy models ($\chi_{r}^{2}$ = 2.3) are similar to the best fitting model, but with no clouds or hazes ($\alpha_{haze}$ = 0.0, $\alpha_{cloud}$ = 0.0) and extreme cloudiness ($\alpha_{cloud}$ = 1.0), respectively. The flat model ($\chi_{r}^{2}$ = 2.7) represents a gray (featureless) spectrum. The models shown here do not predict that Na {\sc i} or K {\sc i} should be present in the transmission spectrum, indicating that these species may be depleted in the atmosphere of HAT-P-32Ab \citep{Burrows99}.  

\subsection{Retrieving HAT-P-32Ab's Atmospheric Properties}
\label{sec:retrieval}


\begin{deluxetable}{cc}
\tabletypesize{\scriptsize}
\tablewidth{0pt}
\tablecolumns{3}
\tablecaption{PLATON atmospheric retrieval results for HAT-P-32Ab \label{tab:platon}}

\tablehead{\colhead{Parameter}  & \colhead{\emph{HST}+\emph{Spitzer}}  }
\startdata 
Planetary radius, $R_{p}$ [$R_{J}$] 				& 1.96$_{-0.00}^{+0.00}$		\\
Isothermal temperature, T [K] 	   					& 1248$_{-92}^{+92}$ 			\\
Metallicity, log(Z)				   					& 2.41$_{-0.07}^{+0.06}$		\\
Carbon-to-oxygen ratio, C/O 		   				& 0.12$_{-0.04}^{+0.08}$		\\
Cloudtop pressure, log($P_{cloud}$ [Pa])			& 3.61$_{-1.03}^{+0.91}$ 	 	\\
Scattering, log(scattering factor)				 	& 1.00$_{-0.28}^{+0.37}$		\\
Scattering slope 									& 9.02$_{-1.00}^{+0.58}$ 		\\
\enddata                                                                     

\end{deluxetable}

\label{tab:platon}

\begin{figure}
    \centering
    \includegraphics[scale=0.60]{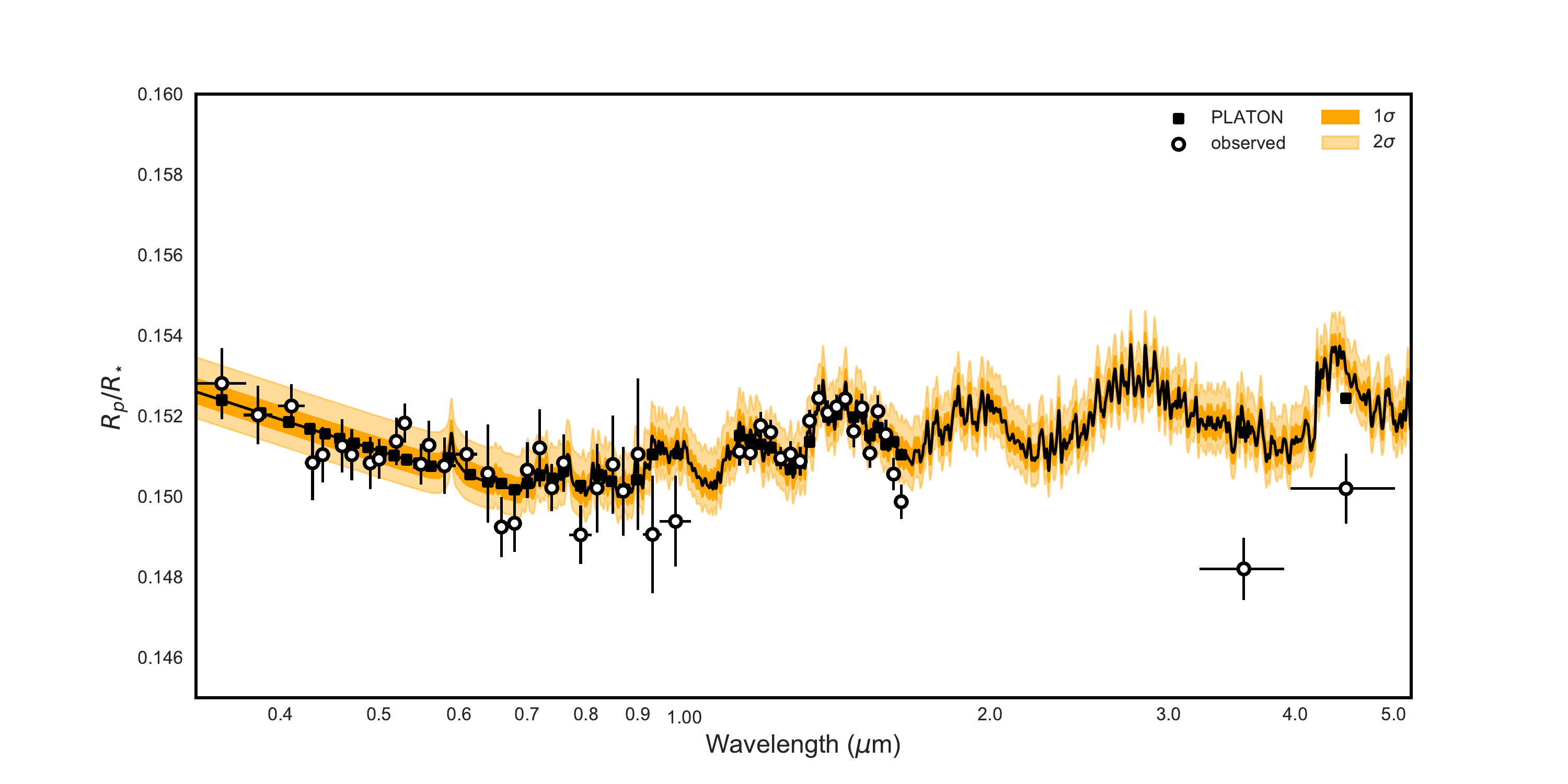}
    \caption{Transmission spectrum of HAT-P-32Ab measured with \textit{HST}+\textit{Spitzer} (open circles). The best-fit model binned to the resolution of the data (squares) and the median fit to the retrieved spectrum (black line) are shown. The shaded regions indicate the 1$\sigma$ (medium orange) and 2$\sigma$ (light orange) credible intervals. } 
    \label{fig:retrieval}
\end{figure}

Although the forward model fits described in \S \ref{sec:fits} well match the red optical and near-infrared portions of the transmission spectrum, the best-fitting model poorly constrains the data in the blue optical. We therefore retrieve the atmospheric properties of our \textit{HST}+\textit{Spitzer} transmission spectrum using the Python-based PLanetary Atmospheric Transmission for Observer Noobs (PLATON)\footnote{\url{https://github.com/ideasrule/platon}} \citep{Zhang19} code to better constrain HAT-P-32Ab's atmosphere\footnote{PLATON has been tested against the ATMO Retrieval Code (ARC, \citealt{Tremblin15}), and both codes have been found to be in agreement \citep{Zhang19}. The computational speed of PLATON introduces some limitations in the accuracy of the results. The opacity sampling method introduces white noise, resulting in spikier retrieved spectra (compared to ATMO) that are accurate to only 100 ppm. To first order, white noise inaccuracies should only affect the width of the posterior distributions \citep{Garland19}. For retrievals of low-resolution transmission spectra such as our \textit{HST}+\textit{Spitzer} observations, however, the intrinsic wavelength spacing of the code largely averages out inaccuracies in the opacity sampling \citep{Zhang19}.}. The results of the full optical to infrared retrieval analysis for this planet are shown in Figure \ref{fig:retrieval} and Table \ref{tab:platon}.



\begin{figure}
    \centering
     \includegraphics[scale=0.48]{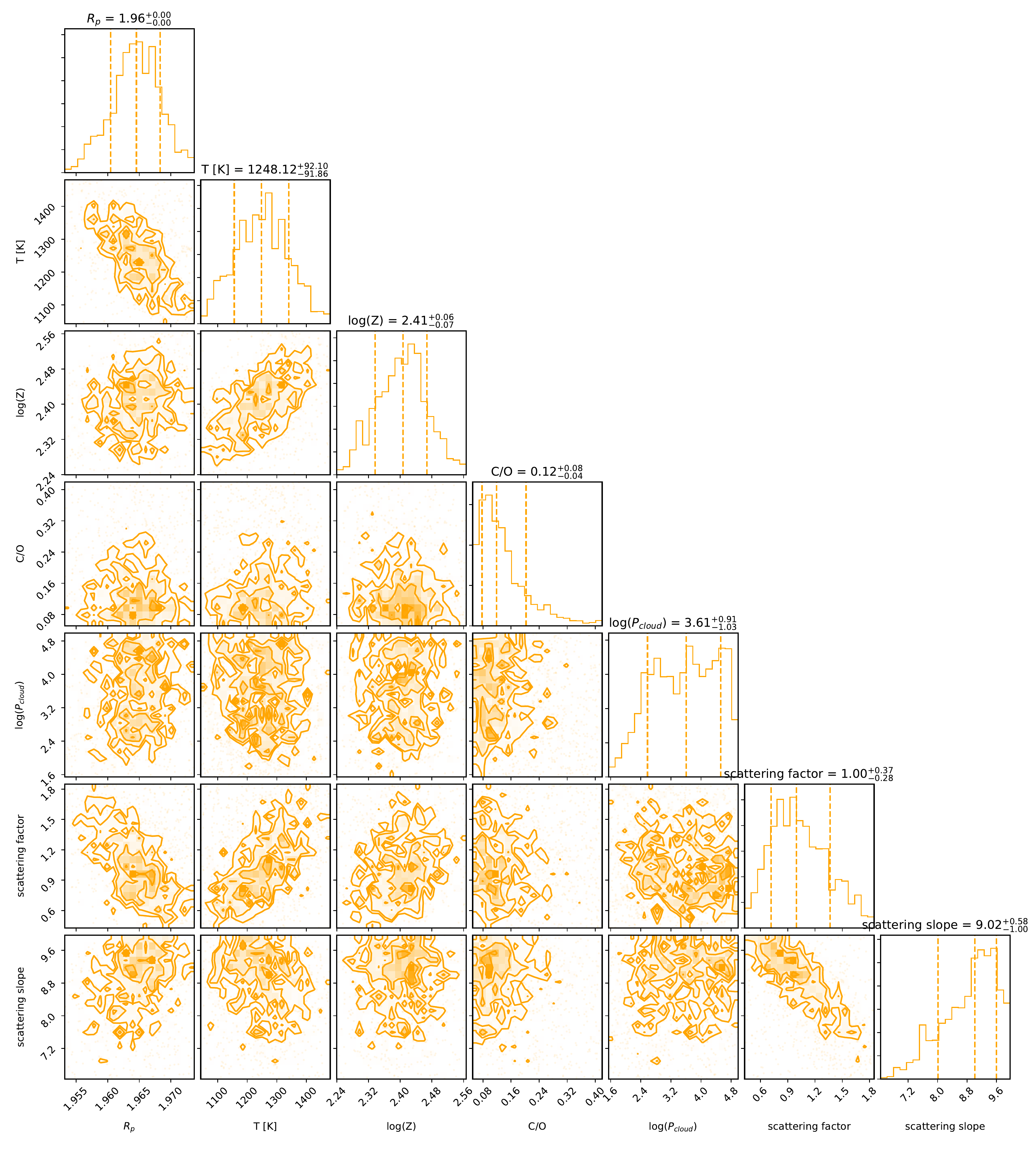}
    \caption{Pairs plot showing distributions of retrieved parameters from the \textit{HST}+\textit{Spitzer} transmission spectrum. We show constraints on the planetary radius, temperature of the isothermal planet atmosphere, metallicity (in solar units), C/O, cloud-top pressure (in Pascals), log(scattering factor), and scattering slope.}
    \label{fig:corner_hst}
\end{figure}

We constrain the planetary radius $R_{p}$, temperature of the isothermal part of the atmosphere $T_{p}$, atmospheric metallicity log($Z$), carbon-to-oxygen ratio C/O, cloud-top pressure $P_{cloud}$,  the factor by which the absorption coefficient is stronger than Rayleigh scattering at the reference wavelength of 1 $\mu$m (log(scattering factor)), and the scattering slope. We use flat priors for $R_{p}$, $T_{p}$, log($Z$), and C/O, with upper and lower bounds for $R_{p}$ and $T_{p}$ from \citet{Tregloan17}. Our metallicity and C/O priors are set by PLATON's pre-computed equilibrium chemistry grid \citep{Zhang19}. Pairs plots showing the distributions of retrieved parameters. 
are presented in Figure \ref{fig:corner_hst}. 
We initially performed our retrievals including activity in our fits (parametrized by spot size and temperature contrast), but found that the model with no stellar heterogeneities was preferred. This finding is consistent with the star appearing quiet in the optical photometry as described in \S \ref{sec:activity}. We therefore adopt the results from the fits without activity henceforth in the paper.

The results of our retrieval fits to the \textit{HST}+\textit{Spitzer} spectrum are summarized in Table \ref{tab:platon}. The best-fit retrieved spectrum is consistent with an isothermal temperature of 1248$_{-92}^{+92}$ K, a thick cloud deck, enhanced Rayleigh scattering, and $\sim$10x H$_{2}$O abundance. The inferred atmospheric metallicity of 2.41$_{-0.07}^{+0.06}$ x solar follows the observed mass-metallicity trend for the Solar System. We also retrieve a sub-solar C/O of 0.12$_{-0.04}^{+0.08}$, a log cloudtop pressure of 3.61$_{-1.03}^{+0.91}$, a scattering factor of 1.00$_{-0.28}^{+0.37}$, and a scattering slope of 9.02$_{-1.00}^{+0.58}$. 


In comparison with the best-fitting ATMO forward model (\S \ref{sec:fits}), we note that the estimated subsolar values for C/O from our ATMO and PLATON fits confirm the presence of clouds in the atmosphere of this planet \citep{Helling19}.  The atmospheric metallicity from ATMO (log($Z$) $\sim$ -0.04; \citealt{Bertelli94}), however, does not well match the constrained PLATON metallicity for the broadband \textit{HST}+\textit{Spitzer} spectrum. 

The retrieved limb temperature from PLATON is lower than the equilibrium temperature of HAT-P-32Ab. This finding is in accordance with other retrieval results from the literature in which retrieved temperatures have been found to be notably cooler ($\sim$200$-$600 K) than planetary equilibrium temperatures (c.f. Table 1 of \citealt{MacDonald20}). These lower retrieved temperatures appear to be the result of applying 1D atmospheric models to planetary spectra with different morning-evening terminator compositions \citep{MacDonald20}. Although 1D retrievals provide an acceptable fit to observations, they artificially shift atmospheric parameters away from terminator-averaged properties. As a result, the retrieved temperature profiles are hundreds of degrees cooler and have weaker temperature gradients than reality. 


Furthermore, our retrieval and forward model fits confirm a cloudy atmosphere for this planet. Our findings also corroborate previous PanCET results for this planet suggesting a Bond albedo of $A_B$ $<$ 0.4 and poor atmospheric re-circulation \citep{Nikolov18}, consistent with the measured geometric albedo of $A_{g}$ $<$ 0.2 for this planet by \citet{Mallonn19}, as well as previous studies showing that planets with higher stellar irradiation levels have greater day-night temperature contrasts and lower re-circulation efficiencies (e.g., \citealt{Schwartz15,Kataria16,Schwartz17}).





\section{HAT-P-32A\MakeLowercase{b} in Context}
\label{sec:context}
\begin{figure}
    \centering
    \includegraphics[scale=0.65]{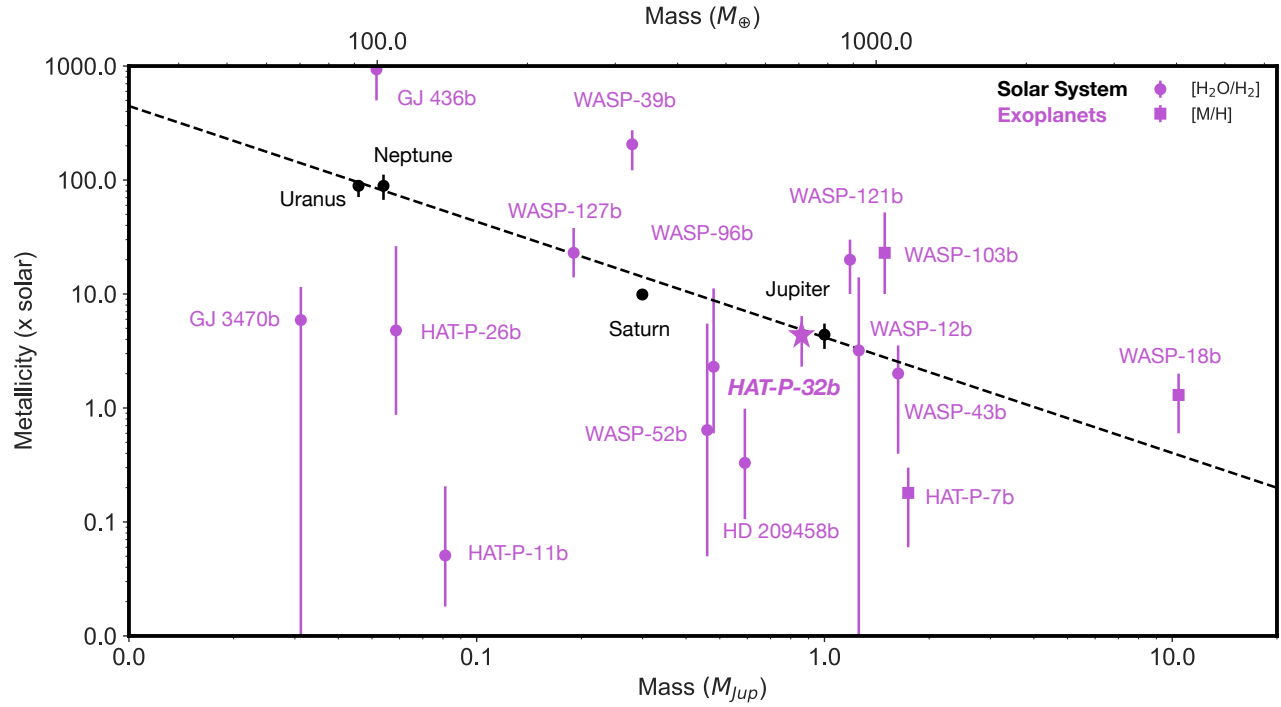}
    \caption{The observed mass-metallicity trend for transiting exoplanets with metallicity constraints from [H$_{2}$O/H$_{2}$] (purple points) or [M/H] (purple squares) and the Solar System gas and ice giants (black points). The dashed black line corresponds to a linear fit in log-log space to the Solar System points.}
    \label{fig:mass_metallicity}
\end{figure}

We interpret the optical to infrared transmission spectrum of HAT-P-32Ab in light of the observed mass-metallicity relation for exoplanets and theoretical predictions for inferring a priori the presence of clouds in exoplanet atmospheres. Our retrieval of the 0.3$-$5.0 $\mu$m \textit{HST}+\textit{Spitzer} spectrum is consistent with the presence of a thick cloud deck, enhanced Rayleigh scattering, and $\sim$10x solar H$_{2}$O abundance. This value is consistent with the H$_{2}$O abundance constraint for HAT-P-32Ab's atmosphere inferred by \citet{Damiano17} using an independent reduction of the WFC3 data set only. Based on the metallicity inferred from PLATON (log($Z/Z_{\odot}$) = 2.41$_{-0.07}^{+0.06}$), we find that HAT-P-32Ab follows the expected mass-metallicity trend for exoplanets based on our Solar System gas giants (e.g., \citealt{Kreidberg14,Wakeford18}). Figure \ref{fig:mass_metallicity} shows HAT-P-32Ab among other exoplanets with metallicity constraints from water abundances (or a sodium abundance constraint in the case of WASP-96b; \citealt{Nikolov18b}), compared to the Solar System gas and ice giants.

Furthermore, the fractional change in atmospheric scale height (H$_{2}$O$-$J) has been suggested as a near-infrared diagnostic for the degree of cloudiness of an exoplanet atmosphere \citep{Stevenson16}. We measure the strength of the water feature using the method of \citet{Stevenson16}, which requires computing the difference in transit depth between the J-band peak (1.36$-$1.44 $\mu$m) and baseline (1.22$-$1.30 $\mu$m) spectral regions and then dividing by the change in transit depth $\Delta D$, which corresponds to a one scale height change in altitude.  $\Delta D$ is given by the relation $\Delta D$ $\sim$ 2$HR_{p}$/$R_{\star}^{2}$, where $H$ is the atmospheric scale height, $R_{p}$ is the planetary radius, and $R_{\star}$ is the stellar radius. $H$ is computed using an equilibrium temperature assuming the planet has zero albedo (i.e., absorbs all incident flux) and consequently re-radiates that energy over its entire surface as a blackbody of that temperature. With a sample of 12, the \citet{Stevenson16} study found that planets with equilibrium temperatures higher than 700 K and surface gravities greater than log($g$) = 2.8 (cgs) are more likely to be cloud-free \citep{Stevenson16}. 

\begin{figure}
    \centering
    \includegraphics[scale=0.65]{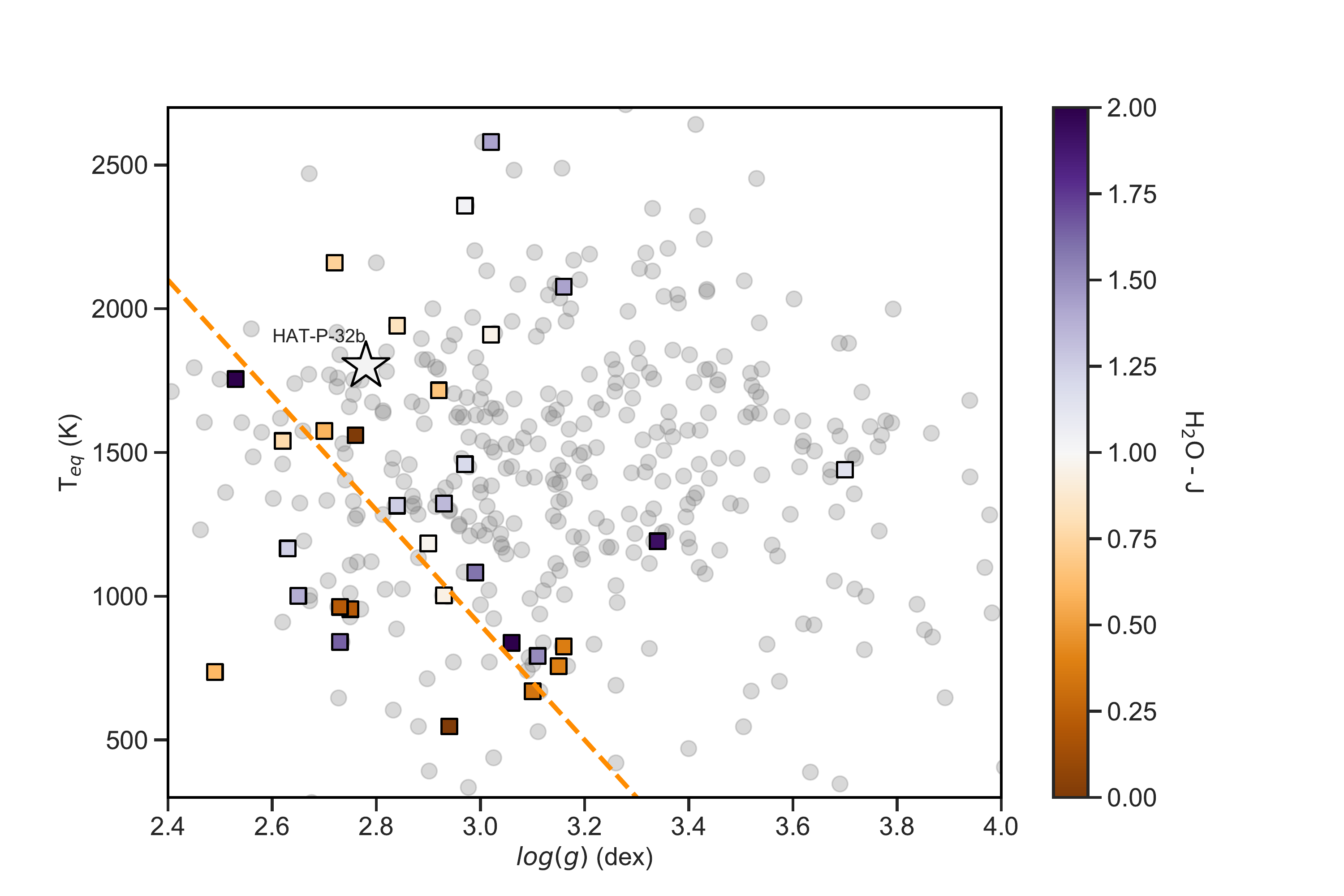}
    \caption{Amplitude of the observed 1.4 $\mu$m H$_{2}$O bandhead as a function of planetary equilibrium temperature and surface gravity (squares), color coded by the strength of the feature. Exoplanets with mass and radius measurements (gray circles) are shown for reference. The dashed orange line shows the proposed divide \citep{Stevenson16} to delineate between cloudy versus clear planets in the $T_{eq}$ $-$ log($g$) phase space. HAT-P-32Ab (white star) crosses this proposed divide and falls in the region theorized to be populated by clear atmosphere planets.}
    \label{fig:logg_Teq}
\end{figure}

We similarly search for trends in cloudiness in the $T_{eq}$ $-$ log($g$) phase space using the expanded sample of 37 planets for which we can measure the H$_{2}$O$-$J index, shown in Figure \ref{fig:logg_Teq}. We use the WFC3 data presented in \citet{Wakeford19}, reduced in a uniformly consistent manner, to compute  H$_{2}$O$-$J. We note that the reductions from \citet{Tsiaras18} also present consistent results. We find that several planets lie along the proposed divide \citep{Stevenson16} to delineate between two classes of cloudy versus clear planets in the $T_{eq}$ $-$ log($g$) phase space. For our more complete sample, the trend is further muddled by the fact that planets such as HAT-P-32Ab with flat transmission spectra indicating the presence of clouds, fall in the region of this parameter space theorized to be populated by cloud-free planets. Moreover, the optical cloudiness index set forth by \citet{Heng16} suggests that more irradiated planets are more likely to be cloud-free. With a planetary temperature constraint of $T_{p}$ = 1801 $\pm$ 18 K \citep{Tregloan17}, HAT-P-32Ab does not fit this prediction as it is a highly irradiated planet with a thick cloud layer. These findings suggest that other physical parameters impact cloud opacities in the atmospheres of close-in giant exoplanets and therefore need to be considered in interpreting atmospheric observations.   

\section{Summary} \label{sec:summary}

We measured the transmission spectrum of the hot Jupiter HAT-P-32Ab over the 0.3$-$5 $\mu$m wavelength range with \textit{HST}+\textit{Spitzer} transit observations. Below we summarize our conclusions about the atmospheric properties of this planet based on these measurements.   

\begin{itemize}

\item The transmission spectrum is characterized by an optical Rayleigh scattering slope, a weak H$_{2}$O feature at 1.4 $\mu$m, and no evidence of alkali absorption features. Compared to a grid of 1D radiative-convective equilibrium models, the best-fitting model indicates the presence of clouds/hazes, consistent with previous ground-based observations (Figure \ref{fig:tr_spec}).

\item We retrieve the planet's atmospheric properties (Figure \ref{fig:retrieval}) using PLATON. The results are consistent with $\sim$10x solar H$_{2}$O abundance and are in agreement with the observed mass-metallicity relation for exoplanets (Figure \ref{fig:mass_metallicity}).  


\item We consider theoretical predictions for inferring a priori the presence of clouds in exoplanet atmospheres (\citealt{Stevenson16,Fu17}). We find that HAT-P-32Ab calls these hypotheses into question, since it is among a handful of planets that cross the proposed divide \citep{Stevenson16} to delineate between two classes of cloudy versus clear exoplanets in the $T_{eq}$ $-$ log($g$) phase space (Figure \ref{fig:logg_Teq}).
 
\end{itemize}

\acknowledgments
M.K.A. thanks Michael Zhang for useful discussions. This paper makes use of observations from the NASA/ESA Hubble Space Telescope, obtained at the Space Telescope Science Institute, which is operated by the Association of Universities for Research in Astronomy, Inc., under NASA contract NAS 5-26555. These observations are associated with \textit{HST} GO programs 14767 and 14260. This work is based on observations made with the Spitzer Space Telescope, which is operated by the Jet Propulsion Laboratory, California Institute of Technology under a contract with NASA. M.K.A. acknowledges support by the National Science Foundation through a Graduate Research Fellowship. V.B. has received funding from the European Research Council (ERC) under the European Union's Horizon 2020 research and innovation program (project Four Aces; grant agreement no. 724427).  J.M.D acknowledges the European Research Council (ERC) European Union’s Horizon 2020 research and innovation programme (grant agreement no. 679633; Exo-Atmos) and the Amsterdam Academic Alliance (AAA) Program. G.W.H. acknowledges long-term support from NASA, NSF, Tennessee State University and the State of Tennessee through its Centers of Excellence Program and from the Space Telescope Science Institute under grant HST-GO-14767. J.S.F. acknnowledges support from the Spanish MINECO through grant AYA2016-79425-C3-2-P. H.R.W. acknowledges support from the Giacconi Prize Fellowship at the Space Telescope Science Institute, which is operated by the Association of Universities for Research in Astronomy, Inc.

\appendix\section{White Light Curve Systematics Model Selection}\label{sec.sys}

As described in \S \ref{sec:stis_lc} and \S \ref{sec:wfc3_lc}, we detrended the \textit{HST} white light curves using a family of systematics models spanning all possible combinations of the detrending parameters for STIS and WFC3 (c.f. Appendix B1 of \citealt{Alam18} for further details). For each of the systematics models used, we performed separate fits for each model and marginalized over the entire set of models, assuming equally weighted priors. Table \ref{tab:systematics} lists the combinations of detrending parameters for the STIS and WFC3 systematics models. For both data sets, the model with the lowest Aikake Information Criterion (AIC) value was selected for detrending. The selection of these models is summarized in Table \ref{tab:systematic_models}.  

\setcounter{table}{0} \renewcommand{\thetable}{A\arabic{table}}


\startlongtable
\begin{deluxetable}{cl}
\tabletypesize{\scriptsize}
\tablewidth{0pt}
\tablecolumns{2}
\tablecaption{\textit{HST} white light curve systematics models  
\label{tab:systematics}}

\tablehead{\colhead{ }  & \colhead{Model}  }
\startdata 
\sidehead{STIS G430L models}
   1 & $\phi_{t}+\phi^{2}_{t}+\phi^{3}_{t}+\phi^{4}_{t}+t$\\
   2 & $\phi_{t}+\phi^{2}_{t}+\phi^{3}_{t}+\phi^{4}_{t}+t+\omega+x^2$\\
   3 & $\phi_{t}+\phi^{2}_{t}+\phi^{3}_{t}+\phi^{4}_{t}+t+x+y^2$\\
   4 & $\phi_{t}+\phi^{2}_{t}+\phi^{3}_{t}+\phi^{4}_{t}+t+x^2+y$\\
   5 & $\phi_{t}+\phi^{2}_{t}+\phi^{3}_{t}+\phi^{4}_{t}+t+\omega$\\
   6 & $\phi_{t}+\phi^{2}_{t}+\phi^{3}_{t}+\phi^{4}_{t}+t+x$\\
   7 & $\phi_{t}+\phi^{2}_{t}+\phi^{3}_{t}+\phi^{4}_{t}+t+\omega+x^2+y$\\
   8 & $\phi_{t}+\phi^{2}_{t}+\phi^{3}_{t}+\phi^{4}_{t}+t+x+x^2+y$\\
   9 & $\phi_{t}+\phi^{2}_{t}+\phi^{3}_{t}+\phi^{4}_{t}+t+y$\\
  10 & $\phi_{t}+\phi^{2}_{t}+\phi^{3}_{t}+\phi^{4}_{t}+t+\omega+x$\\
  11 & $\phi_{t}+\phi^{2}_{t}+\phi^{3}_{t}+\phi^{4}_{t}+t+\omega+y$\\
  12 & $\phi_{t}+\phi^{2}_{t}+\phi^{3}_{t}+\phi^{4}_{t}+t+\omega+x+y$\\
  13 & $\phi_{t}+\phi^{2}_{t}+\phi^{3}_{t}+\phi^{4}_{t}+t+\omega+\omega^2+y$\\
  14 & $\phi_{t}+\phi^{2}_{t}+\phi^{3}_{t}+\phi^{4}_{t}+t+\omega+x+x^2$\\
  15 & $\phi_{t}+\phi^{2}_{t}+\phi^{3}_{t}+\phi^{4}_{t}+t+\omega+y+y^2$\\
  16 & $\phi_{t}+\phi^{2}_{t}+\phi^{3}_{t}+\phi^{4}_{t}+t+x+y$\\
  17 & $\phi_{t}+\phi^{2}_{t}+\phi^{3}_{t}+\phi^{4}_{t}+t+\omega+\omega^2+x$\\
  18 & $\phi_{t}+\phi^{2}_{t}+\phi^{3}_{t}+\phi^{4}_{t}+t+x+x^2$\\
  19 & $\phi_{t}+\phi^{2}_{t}+\phi^{3}_{t}+\phi^{4}_{t}+t+x+x^2+y$\\
  20 & $\phi_{t}+\phi^{2}_{t}+\phi^{3}_{t}+\phi^{4}_{t}+t+y+y^2$\\
  21 & $\phi_{t}+\phi^{2}_{t}+\phi^{3}_{t}+\phi^{4}_{t}+t+y$\\
  22 & $\phi_{t}+\phi^{2}_{t}+\phi^{3}_{t}+\phi^{4}_{t}+t+\omega+\omega^2+x$\\
  23 & $\phi_{t}+\phi^{2}_{t}+\phi^{3}_{t}+\phi^{4}_{t}+t+\omega+\omega^2+x+y$\\
  24 & $\phi_{t}+\phi^{2}_{t}+\phi^{3}_{t}+\phi^{4}_{t}+t+\omega+\omega^2+x+x^2+y+y^2$\\
  25 & $\phi_{t}+\phi^{2}_{t}+\phi^{3}_{t}+\phi^{4}_{t}+t+\omega+\omega^2+x+x^2+y^2$\\
\sidehead{STIS G750L models}
   1 & $\phi_{t}+\phi^{2}_{t}+\phi^{3}_{t}+\phi^{4}_{t}+t$\\
   2 & $\phi_{t}+\phi^{2}_{t}+\phi^{3}_{t}+\phi^{4}_{t}+t+\omega+x$\\
   3 & $\phi_{t}+\phi^{2}_{t}+\phi^{3}_{t}+\phi^{4}_{t}+t+x+y^2$\\
   4 & $\phi_{t}+\phi^{2}_{t}+\phi^{3}_{t}+\phi^{4}_{t}+t+\omega^2$\\
   5 & $\phi_{t}+\phi^{2}_{t}+\phi^{3}_{t}+\phi^{4}_{t}+t+\omega$\\
   6 & $\phi_{t}+\phi^{2}_{t}+\phi^{3}_{t}+\phi^{4}_{t}+t+x$\\
   7 & $\phi_{t}+\phi^{2}_{t}+\phi^{3}_{t}+\phi^{4}_{t}+t+\omega+x^2+y^3$\\
   8 & $\phi_{t}+\phi^{2}_{t}+\phi^{3}_{t}+\phi^{4}_{t}+t+x+y$\\
   9 & $\phi_{t}+\phi^{2}_{t}+\phi^{3}_{t}+\phi^{4}_{t}+t+y$\\
  10 & $\phi_{t}+\phi^{2}_{t}+\phi^{3}_{t}+\phi^{4}_{t}+t+\omega+x$\\
  11 & $\phi_{t}+\phi^{2}_{t}+\phi^{3}_{t}+\phi^{4}_{t}+t+\omega+y$\\
  12 & $\phi_{t}+\phi^{2}_{t}+\phi^{3}_{t}+\phi^{4}_{t}+t+\omega+x+y$\\
  13 & $\phi_{t}+\phi^{2}_{t}+\phi^{3}_{t}+\phi^{4}_{t}+t+y$\\
  14 & $\phi_{t}+\phi^{2}_{t}+\phi^{3}_{t}+\phi^{4}_{t}+t+\omega+x^2$\\
  15 & $\phi_{t}+\phi^{2}_{t}+\phi^{3}_{t}+\phi^{4}_{t}+t+\omega+y+y^2$\\
  16 & $\phi_{t}+\phi^{2}_{t}+\phi^{3}_{t}+\phi^{4}_{t}+t+\omega+y+y^2+y^3$\\
  17 & $\phi_{t}+\phi^{2}_{t}+\phi^{3}_{t}+\phi^{4}_{t}+t+\omega+\omega^2+x$\\
  18 & $\phi_{t}+\phi^{2}_{t}+\phi^{3}_{t}+\phi^{4}_{t}+t+x+y^2$\\
  19 & $\phi_{t}+\phi^{2}_{t}+\phi^{3}_{t}+\phi^{4}_{t}+t+x$\\
  20 & $\phi_{t}+\phi^{2}_{t}+\phi^{3}_{t}+\phi^{4}_{t}+t+y+y^2$\\
  21 & $\phi_{t}+\phi^{2}_{t}+\phi^{3}_{t}+\phi^{4}_{t}+t+y+y^2+y^3$\\
  22 & $\phi_{t}+\phi^{2}_{t}+\phi^{3}_{t}+\phi^{4}_{t}+t+\omega+\omega^2+x$\\
  23 & $\phi_{t}+\phi^{2}_{t}+\phi^{3}_{t}+\phi^{4}_{t}+t+\omega+\omega^2+x+y$\\
  24 & $\phi_{t}+\phi^{2}_{t}+\phi^{3}_{t}+\phi^{4}_{t}+t+\omega+\omega^2+x+y$\\
  25 & $\phi_{t}+\phi^{2}_{t}+\phi^{3}_{t}+\phi^{4}_{t}+t+\omega+\omega^2+x^2$\\
\sidehead{WFC3 G141 models}
   1 & $a_{hst2}+a_{hst3}+a_{hst4}+a_{x1}+a_{x2}+a_{x3}+a_{y1}+a_{y2}+a_{y3}+a_{t1}$\\
   2 & $a_{hst3}+a_{hst4}+a_{x1}+a_{x2}+a_{x3}+a_{y1}+a_{y2}+a_{y3}+a_{t1}$\\
   3 & $a_{hst4}+a_{x1}+a_{x2}+a_{x3}+a_{y1}+a_{y2}+a_{y3}+a_{t1}$\\
   4 & $a_{x1}+a_{x2}+a_{x3}+a_{y1}+a_{y2}+a_{y3}+a_{t1}$\\
   5 & $a_{hst3}+a_{x1}+a_{x2}+a_{x3}+a_{y1}+a_{y2}+a_{y3}+a_{t1}$\\
   6 & $a_{hst2}+a_{hst4}+a_{x1}+a_{x2}+a_{x3}+a_{y1}+a_{y2}+a_{y3}+a_{t1}$\\
   7 & $a_{hst2}+a_{x1}+a_{x2}+a_{x3}+a_{y1}+a_{y2}+a_{y3}+a_{t1}$\\
   8 & $a_{hst2}+a_{hst3}+a_{x1}+a_{x2}+a_{x3}+a_{y1}+a_{y2}+a_{y3}+a_{t1}$\\
   9 & $a_{hst3}+a_{hst4}+a_{x1}+a_{x2}+a_{x3}+a_{y1}+a_{y2}+a_{y3}+a_{t1}$\\
  10 & $a_{hst4}+a_{x1}+a_{x2}+a_{x3}+a_{y1}+a_{y2}+a_{y3}+a_{t1}$\\
  11 & $a_{x1}+a_{x2}+a_{x3}+a_{y1}+a_{y2}+a_{y3}+a_{t1}$\\
  12 & $a_{hst3}+a_{x1}+a_{x2}+a_{x3}+a_{y1}+a_{y2}+a_{y3}+a_{t1}$\\
  13 & $a_{hst2}+a_{hst4}+a_{x1}+a_{x2}+a_{x3}+a_{y1}+a_{y2}+a_{y3}+a_{t1}$\\
  14 & $a_{hst2}+a_{x1}+a_{x2}+a_{x3}+a_{y1}+a_{y2}+a_{y3}+a_{t1}$\\
\enddata
\end{deluxetable}
\label{tab.systematics}


\startlongtable
\begin{deluxetable*}{cccc}
\tabletypesize{\scriptsize}
\tablewidth{0pt}
\tablecolumns{4}
\tablecaption{Systematics model selection for the STIS \& WFC3 white light curves
\label{tab:systematic_models}}

\tablehead{\colhead{Model}  & \colhead{$\chi_{r}^{2}$} & \colhead{AIC} & \colhead{d.o.f} } 
\startdata 
\sidehead{STIS G430L (visit 72)}
1  & 1.75 & 65.07 & 28  \\ 
2  & 1.60 & 61.63 & 26  \\ 
3  & 1.62 & 62.21 & 26  \\ 
4  & 1.76 & 65.76 & 26  \\ 
5  & 1.54 & 59.68 & 27  \\ 
6  & 1.57 & 60.41 & 27  \\ 
7  & 1.73 & 64.78 & 27  \\ 
8  & 1.41 & 57.24 & 25  \\ 
9  & 1.58 & 61.01 & 26  \\ 
10 & 1.55 & 60.84 & 25  \\ 
11 & 1.60 & 61.63 & 26  \\ 
12 & 1.40 & 57.00 & 25  \\ 
13 & 1.63 & 62.87 & 25  \\ 
14 & 1.58 & 61.95 & 24  \\ 
15 & 1.63 & 62.82 & 25  \\ 
16 & 1.55 & 61.16 & 24  \\ 
17 & 1.44 & 58.66 & 24  \\ 
18 & 1.62 & 62.21 & 26  \\ 
19 & 1.67 & 62.83 & 25  \\ 
20 & 1.78 & 66.24 & 26  \\ 
21 & 1.53 & 60.30 & 25  \\ 
22 & 1.55 & 60.65 & 25  \\ 
23 & 1.59 & 62.27 & 24  \\ 
24 & 1.70 & 65.39 & 22  \\ 
25 & 1.56 & 63.24 & 20  \\ 
\sidehead{STIS G430L (visit 73)}
 1 & 2.76 & 90.53 & 27 \\ 
 2 & 2.75 & 88.77 & 25 \\ 
 3 & 1.79 & 64.84 & 25 \\ 
 4 & 2.08 & 52.03 & 25 \\ 
 5 & 2.78 & 90.77 & 26 \\ 
 6 & 2.43 & 81.39 & 26 \\ 
 7 & 2.06 & 71.57 & 26 \\ 
 8 & 2.52 & 82.54 & 24 \\ 
 9 & 2.13 & 73.24 & 25 \\ 
10 & 2.11 & 72.54 & 24 \\ 
11 & 2.88 & 91.95 & 25 \\ 
12 & 2.99 & 93.95 & 24 \\ 
13 & 2.11 & 72.63 & 24 \\ 
14 & 2.60 & 83.85 & 23 \\ 
15 & 1.65 & 61.59 & 24 \\ 
16 & 1.72 & 63.46 & 23 \\ 
17 & 2.56 & 82.81 & 23 \\ 
18 & 2.49 & 82.24 & 25 \\ 
19 & 2.57 & 83.61 & 24 \\ 
20 & 1.58 & 59.59 & 25 \\ 
21 & 1.64 & 61.46 & 24 \\ 
22 & 2.52 & 82.50 & 24 \\ 
23 & 2.13 & 73.02 & 23 \\ 
24 & 1.70 & 63.63 & 21 \\ 
25 & 1.79 & 66.00 & 19 \\ 
\sidehead{STIS G750L (visit 74)}
 1 & 1.99 & 57.10 & 27 \\ 	
 2 & 1.74 & 50.82 & 25 \\ 	
 3 & 2.06 & 59.96 & 25 \\ 	
 4 & 1.62 & 53.93 & 25 \\ 	
 5 & 2.03 & 53.82 & 26 \\ 	
 6 & 1.99 & 59.03 & 26 \\   
 7 & 1.73 & 56.01 & 26 \\ 	
 8 & 1.99 & 55.67 & 25 \\ 	
 9 & 1.76 & 53.18 & 25 \\ 	
10 & 1.76 & 54.84 & 24 \\ 	
11 & 1.97 & 55.68 & 25 \\ 	
12 & 2.03 & 53.82 & 26 \\ 	
13 & 1.69 & 54.83 & 24 \\ 	
14 & 1.72 & 46.94 & 23 \\ 	
15 & 1.83 & 54.71 & 24 \\ 	
16 & 1.89 & 56.18 & 23 \\ 	
17 & 2.03 & 57.10 & 24 \\ 	
18 & 1.79 & 53.42 & 25 \\ 	
19 & 1.70 & 47.31 & 24 \\ 	
20 & 1.80 & 57.61 & 25 \\ 	
21 & 1.85 & 59.32 & 24 \\ 	
22 & 2.03 & 57.10 & 24 \\ 	
23 & 1.82 & 56.59 & 23 \\ 	
24 & 1.83 & 51.60 & 21 \\ 	
25 & 1.67 & 61.00 & 21 \\ 	
\sidehead{WFC3 G141 visit 01}
 1 & 1.08 & 80.20   & 52 \\ 
 2 & 1.07 & 78.62   & 53 \\ 
 3 & 2.10 & 133.51  & 54 \\ 
 4 & 2.94 & 179.86  & 55 \\ 
 5 & 1.44 & 97.54   & 54 \\ 
 6 & 1.99 & 127.59  & 53 \\ 
 7 & 2.39 & 149.55  & 54 \\ 
 8 & 1.14 & 82.38   & 53 \\ 
 9 & 1.04 & 78.09   & 52 \\ 
10 & 1.08 & 79.32   & 53 \\ 
11 & 1.66 & 109.82  & 54 \\ 
12 & 1.46 & 99.28   & 53 \\ 
13 & 1.07 & 79.41   & 52 \\ 
14 & 1.23 & 87.09   & 53 \\ 
\enddata
\end{deluxetable*}

\label{tab.systematic_models}


\bibliographystyle{yahapj}
\bibliography{refs}

\newpage

\end{document}